\documentclass[11pt]{article}
\usepackage[totalwidth=470pt, totalheight=590pt, centering]{geometry}

\usepackage{amsmath,amssymb,mathrsfs}
\usepackage{bbm} 

\usepackage{graphicx}
\usepackage{amsthm}

\usepackage[dvips,bookmarks=true,hyperfootnotes=false]{hyperref}
\hypersetup{pdfstartview=FitH,pdfhighlight=/O,colorlinks=false}

\newcommand{\margin}[1]{}

\newcommand{\mc}[1]{\mathcal{#1}}

\newcommand{\mbb}[1]{\mathbb{#1}}

\newcommand{\eps}{\varepsilon}

\newcommand{\mdots}{,.\,.\,,}

\newcommand{\dd}{\text{d}}
\long\def\symbolfootnote[#1]#2{\begingroup\def\thefootnote{\fnsymbol{footnote}}
\footnote[#1]{#2}\endgroup}

\title{\LARGE{\textbf{LQG propagator from the new spin foams}}}

\author{Eugenio Bianchi{\it ${\,}^{a}$}, Elena Magliaro{\it ${\,}^{ab}$}, Claudio Perini{\it ${\,}^{ac}$}\\[.35em]
\small{\textit{${}^a$Centre de Physique Th\'eorique de Luminy}}\footnote{Unit\'e mixte de recherche (UMR 6207) du CNRS et des Universit\'es
de Provence (Aix-Marseille I), de la M\'editerran\'ee (Aix-Marseille II) et du Sud (Toulon-Var); laboratoire affili\'e \`a la FRUMAM (FR 2291).}\small{\textit{, case 907, F-13288 Marseille, EU}}\\
\small{\textit{${}^b$Dipartimento di Fisica, Universit\`a degli Studi Roma Tre, I-00146 Roma, EU}}\\
\small{\textit{${}^c$Dipartimento di Matematica, Universit\`a degli Studi Roma Tre, I-00146 Roma, EU}}
}

\date{
\small May 25, 2009}

\begin{document}

\maketitle

\begin{abstract}
We compute metric correlations in loop quantum gravity with the dynamics defined by the \emph{new} spin foam models. The analysis is done at the lowest order in a vertex expansion and at the leading order in a large spin expansion. The result is compared to the graviton propagator of perturbative quantum gravity.
\end{abstract}



\section{Introduction}\label{sec:intro}
In this paper we compute metric correlations in Loop Quantum Gravity (LQG) \cite{Rovelli:2004tv,Thiemann:2007zz,Ashtekar:2004eh} and we compare them with the scaling and the tensorial structure of the graviton propagator in perturbative Quantum Gravity \cite{Veltman:1975vx,Donoghue:1994dn,Burgess:2003jk}. The strategy is the one introduced in \cite{Rovelli:2005yj} and developed in \cite{Bianchi:2006uf,Livine:2006it,Bianchi:2007vf,Christensen:2007rv,Alesci:2007tx,Alesci:2007tg,Alesci:2008ff,Speziale:2008uw}. In particular, we use the boundary amplitude formalism \cite{Rovelli:2004tv,Oeckl:2003vu,Oeckl:2005bv,Conrady:2003en}. The dynamics is implemented in terms of (the group field theory expansion of) the new spin foam models introduced by Engle, Pereira, Rovelli and Livine (EPRL$_\gamma$ model) \cite{Engle:2007wy} and by Freidel and Krasnov  (FK$_\gamma$ model) \cite{Freidel:2007py}. We restrict attention to Euclidean signature and Immirzi parameter smaller than one: $0<\gamma<1$. In this case the two models coincide.

Previous attempts to derive the graviton propagator from LQG adopted the Barrett-Crane spin foam vertex \cite{Barrett:1997gw} as model for the dynamics \cite{Rovelli:2005yj,Bianchi:2006uf,Livine:2006it,Bianchi:2007vf,Christensen:2007rv,Alesci:2007tx,Alesci:2007tg,Alesci:2008ff,Speziale:2008uw} (see also \cite{Speziale:2005ma,Livine:2006ab,Bonzom:2008xd}
 for investigations in the three-dimensional case). The analysis of \cite{Alesci:2007tx,Alesci:2007tg} shows that the Barrett-Crane model fails to give the correct scaling behavior for off-diagonal components of the graviton propagator. The problem can be traced back to a missing coherent cancellation of phases between the intertwiner wave function of the semiclassical boundary state and the intertwiner dependence of the model. The attempt to correct this problem was part of the motivation for the lively search of \emph{new} spin foam models with non-trivial intertwiner dependence \cite{Engle:2007uq,Engle:2007qf,Engle:2007wy,Freidel:2007py,Livine:2007vk}. The intertwiner dynamics of the new models was investigated numerically in \cite{Magliaro:2007nc,Alesci:2008ec,Khavkine:2008kk,Khavkine:2008nb}. The analysis of the large spin asymptotics of the vertex amplitude of the new models was performed in \cite{Conrady:2008ea,Conrady:2008mk,Conrady:2009px} and in \cite{Barrett:2009gg}. In \cite{Alesci:2008ff}, the obstacle that prevented the Barrett-Crane model from yielding the correct behaviour of the propagator was shown to be absent for the new models: the new spin foams feature the correct dependence on intertwiners to allow a coherent cancellation of phases with the boundary semiclassical state. In this paper we restart from scratch the calculation and derive the graviton propagator from the new spin foam models.

In this introduction we briefly describe the quantity we want to compute. We consider a manifold $\mc{R}$ with the topology of a $4$-ball. Its boundary is a $3$-manifold $\Sigma$ with the topology of a $3$-sphere $S^3$. We associate to $\Sigma$ a boundary Hilbert space of states: the LQG Hilbert space $\mc{H}_\Sigma$ spanned by (abstract) spin networks. We call $|\Psi\rangle$ a generic state in $\mc{H}_\Sigma$. A spin foam model for the region $\mc{R}$ provides a map from the boundary Hilbert space to $\mbb{C}$.  We call this map $\langle W|$. It provides a sum over the bulk geometries with a weight that defines our model for quantum gravity. The dynamical expectation value of an operator $\mc{O}$ on the state $|\Psi\rangle$ is defined via the following expression\footnote{This expression corresponds to the standard definition in (perturbative) quantum field theory where the \emph{vacuum} expectation value of a product of local observables is defined as
\begin{equation}
\langle O(x_1)\cdots O(x_n)\rangle_0=\frac{\displaystyle \int D[\varphi] O(x_1)\cdots O(x_n) e^{i S[\varphi]}}{\displaystyle \int D[\varphi]  e^{i S[\varphi]}}\equiv \frac{\displaystyle \int D[\phi] W[\phi] O(x_1)\cdots O(x_n) \Psi_0[\phi]}{\displaystyle \int D[\phi] W[\phi] \Psi_0[\phi]}\;.
\end{equation}
The vacuum state $\Psi_0[\phi]$ codes the boundary conditions at infinity.}
\begin{equation}
\langle \mc{O}\rangle=\frac{\langle W| \mc{O}|\Psi\rangle}{\langle W|\Psi\rangle}\;.
\end{equation}
The operator $\mc{O}$ can be a geometric operator as the area, the volume or the length \cite{Rovelli:1994ge,Ashtekar:1996eg,Ashtekar:1997fb,Ashtekar:1998ak,Major:1999mc,Thiemann:1996at,Bianchi:2008es}. The geometric operator we are interested in here is the (density-two inverse-) metric operator $q^{ab}(x)=\delta^{ij}E^a_i(x) E^b_j(x)$. We focus on the \emph{connected} two-point correlation function $G^{abcd}(x,y)$ on a semiclassical boundary state $|\Psi_0\rangle$. It is defined as
 \begin{equation}
G^{abcd}(x,y)=\langle q^{ab}(x)\; q^{cd}(y)\rangle - \langle q^{ab}(x)\rangle\, \langle q^{cd}(y)\rangle\;.
\label{eq:G}
\end{equation}
The boundary state $|\Psi_0\rangle$  is semiclassical in the following sense: it is peaked on a given configuration of the intrinsic and the extrinsic geometry of the boundary manifold $\Sigma$. In terms of Ashtekar-Barbero variables these boundary data correspond to a couple $(E_0,A_0)$. The boundary data are chosen so that there is a solution of Einstein equations in the bulk which induces them on the boundary. A spin foam model has good semiclassical properties if the dominant contribution to the amplitude $\langle W|\Psi_0\rangle$ comes from the bulk configurations close to the classical $4$-geometries compatible with the boundary data $(E_0,A_0)$. By \emph{classical} we mean that they satisfy Einstein equations. 

The classical bulk configuration we focus on is flat space. The boundary configuration that we consider is the following: we decompose the boundary manifold $S^3$ in five tetrahedral regions with the same connectivity as the boundary of a $4$-simplex; then we choose the intrinsic and the extrinsic geometry to be the ones proper of the boundary of a Euclidean $4$-simplex. By construction, these boundary data are compatible with flat space being a classical solution in the bulk.

For our choice of boundary configuration, the dominant contribution to the amplitude $\langle W|\Psi_0\rangle$ is required to come from bulk configurations close to flat space. The connected two-point correlation function $G^{abcd}(x,y)$ probes the fluctuations of the geometry around the classical configuration given by flat space. As a result it can be compared to the graviton propagator computed in perturbative quantum gravity.

The plan of the paper is the following: in section \ref{sec:boundary state} we introduce the metric operator and construct a semiclassical boundary state; in section \ref{sec:W} we recall the form of the new spin foam models; in section \ref{sec:integral formula} we define the LQG propagator and provide an integral formula for it at the lowest order in a vertex expansion; in section \ref{sec:stationary phase} we compute its large spin asymptotics; in section \ref{sec:expectation values} we discuss expectation values of metric operators; in section \ref{sec:propagator} we present our main result: the scaling and the tensorial structure of the LQG propagator at the leading order of our expansion; in section \ref{sec:perturbative} we attempt a comparison with the graviton propagator of perturbative quantum gravity.

\section{Semiclassical boundary state and the metric operator}\label{sec:boundary state}
Semiclassical boundary states are a key ingredient in the definition of boundary amplitudes. Here we describe in detail the construction of a boundary state peaked on the intrinsic and the extrinsic geometry of the boundary of a Euclidean $4$-simplex. The construction is new: it uses the coherent intertwiners of Livine and Speziale \cite{Livine:2007vk} (see also \cite{Conrady:2009px}) together with a superposition over spins as done in \cite{Rovelli:2005yj,Bianchi:2006uf}. It can be considered as an improvement of the boundary state used in \cite{Alesci:2007tx,Alesci:2007tg,Alesci:2008ff} where Rovelli-Speziale gaussian states \cite{Rovelli:2006fw} for intertwiners were used.

We consider a simplicial decomposition $\Delta_5$ of $S^3$. The decomposition $\Delta_5$ is homeomorphic to the boundary of a $4$-simplex: it consists of five cells $t_a$ which meet at ten faces $f_{ab}$ ($a,b=1\mdots 5$ and $a<b$). Then we consider the sector of the Hilbert space $\mc{H}_\Sigma$ spanned by spin network states with graph $\Gamma_5$ dual to the decomposition $\Delta_5$,
\begin{equation}
\Gamma_5=\;\;\parbox[c]{130pt}{\includegraphics{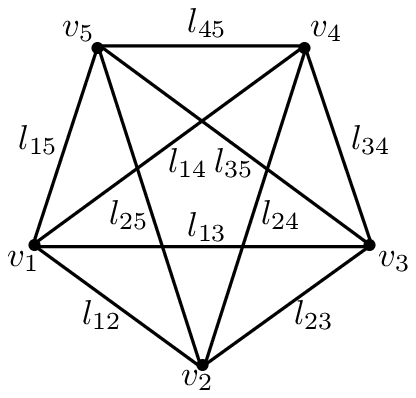}}\;.
\label{eq:G5}
\end{equation}
$\Gamma_5$ is a complete graph with five nodes. We call $v_a$ its nodes and $l_{ab}$ ($a<b$) its ten links. Spin network states supported on this graph are labelled by ten spins $j_{ab}$ ($a<b$) and five intertwiners $i_a$. We denote them by $|\Gamma_5,j_{ab},i_a\rangle$ and call $\mc{H}_{\Gamma_5}$ the Hilbert space they span.
On $\mc{H}_{\Gamma_5}$ we can introduce a \emph{metric operator} smearing the electric field on surfaces dual to links, i.e. considering scalar products of fluxes. We focus on the node $n$ and consider a surface $f_{na}$ which cuts the link from the node $n$ to the node $a$. The flux operator through the surface $f_{na}$, parallel transported in the node $n$, is denoted\footnote{Throughout the paper  $i,j,k\ldots=1,2,3$ are indices for vectors in $\mbb{R}^3$.} $(E_n^a)_i$. It has the following three non-trivial properties:\\
(i) the flux operators $(E_n^a)_i$ and $(E_a^n)_i$ are related by a $SU(2)$ parallel transport $g_{an}$ from the node $a$ to the node $n$ together with a change of sign which takes into account the different orientation of the face $f_{an}$,
\begin{equation}
(E_n^{a})_i=-{(R_{an})_i}^{\,j}\, (E_a^{n})_j  \; ,
\end{equation}
where $R_{an}$ is the rotation which corresponds to the group element $g_{an}$ associated to the link $l_{an}$, i.e. $R_{an}=D^{(1)}(g_{ab})$;\\
(ii) the commutator of two flux operators for the same face $f_{na}$ is\footnote{Throughout the paper we put $c=\hbar=G_\text{Newton}=1$.}
\begin{equation}
[\,(E_n^{a})_i\,,\,(E_n^{a})_j\,]=i\gamma {\eps_{ij}}^k \,(E_n^{a})_k\; ;
\end{equation}
(iii) a spin network state is annihilated by the sum of the flux operators over the faces bounding a node
\begin{equation}\label{gaugeinvariance}
\sum_{c\neq n}(E_n^{c})_i |\Gamma_5,j_{ab},i_a\rangle=0\;.
\end{equation}
This last property follows from the $SU(2)$ gauge invariance of the spin network node.

Using the flux operator we can introduce the density-two inverse-metric operator at the node $n$, projected in the directions normal to the faces $f_{na}$ and $f_{nb}$. It is defined as  $E_n^{a}\cdot E_n^{b}=\delta^{ij}(E_n^{a})_i (E_n^{b})_j$. Its diagonal components $E_n^{a}\!\cdot\! E_n^{a}$ measure the area square of the face $f_{na}$,
\begin{equation}
E_n^{a}\!\cdot\! E_n^{a}\, |\Gamma_5,j_{ab},i_a\rangle= \Big(\gamma\sqrt{j_{na}(j_{na}+1)}\Big)^2\,|\Gamma_5,j_{ab},i_a\rangle\;.
\end{equation}
Spin network states are eigenstates of the diagonal components of the metric operator. On the other hand, the off-diagonal components $E_n^{a}\!\cdot\! E_n^{b}$ with $a\neq b$ measure the dihedral angle between the faces $f_{na}$ and $f_{nb}$ (weighted with their areas). It reproduces the angle operator \cite{Major:1999mc}. Using the recoupling basis for intertwiner space, we have that in general the off-diagonal components of the metric operator have non-trivial matrix elements 
\begin{equation}
E_n^{a}\!\cdot\! E_n^{b}\, |\Gamma_5,j_{ab},i_a\rangle= \sum_{i'_c} {\big(E_n^{a}\!\cdot\! E_n^{b}\big)_{i_c}}^{i'_c}\;|\Gamma_5,j_{ab},i'_a\rangle\;.
\end{equation}
We refer to \cite{Alesci:2007tx,Alesci:2007tg} for a detailed discussion. In particular, from property (ii), we have that some off-diagonal components of the metric operator at a node do not commute \cite{Ashtekar:1998ak}
\begin{equation}
[\,E_n^{a}\!\cdot\! E_n^{b}\,,\,E_n^{a}\!\cdot\! E_n^{c}\,]\neq 0\;.
\end{equation}
From this non-commutativity an Heisenberg inequality for dispersions of metric operators follows. Here we are interested in states which are peaked on a given value of \emph{all} the off-diagonal components of the metric operator and which have dispersion of the order of Heisenberg's bound. Such states can be introduced using the technique of coherent intertwiners \cite{Livine:2007vk,Conrady:2009px}. A coherent intertwiner between the representations $j_1\mdots j_4$ is defined as\footnote{The state $|j,\vec{n}\rangle$ is a spin coherent state. It is labelled by a unit vector $\vec{n}$ or equivalently by a point on the unit sphere. Given a $SU(2)$ transformation $g$ which acts on the vector $+\vec{e}_z$ sending it to the vector $\vec{n}=R e_z$, a spin coherent state is given by $|j,\vec{n}\rangle=D^{(j)}(g) |j,+j\rangle$. As a result, it is defined up to a phase $e^{i \alpha j}$ corresponding to a transformation $\exp( i \alpha  \vec{n}\cdot\vec{J})|j,\vec{n}\rangle=e^{i \alpha j} |j,\vec{n}\rangle$. A phase ambiguity in the definition of the coherent intertwiner (\ref{eq:coherent intertwiner}) follows. Such ambiguity becomes observable when a superposition over $j$ is considered. }
\begin{equation}
\Phi^{m_1\cdots m_4}(\vec{n}_1\mdots \vec{n}_4)=\frac{1}{\sqrt{\Omega(\vec{n}_1\mdots \vec{n}_4)}}\int_{SU(2)}dh \; \prod_{a=1}^{4}\,\langle j_a,m_a| D^{(j_a)}(h) |j_a,\vec{n}_a\rangle
\label{eq:coherent intertwiner}
\end{equation}
and is labelled by four unit vectors $\vec{n}_1\mdots \vec{n}_4$ satisfying the closure condition
\begin{equation}
j_1 \vec{n}_1+\cdots+j_4 \vec{n}_4=0\;.
\label{eq:closure}
\end{equation}
The function $\Omega(\vec{n}_1\mdots \vec{n}_4)$ provides normalization to one of the intertwiner. The function $\Phi^{m_1\cdots m_4}$ is invariant under rotations of the four vectors $\vec{n}_1\mdots\vec{n}_4$. In the following we always assume that this invariance has been fixed with a given choice of orientation\footnote{\label{footnote:orientation}For instance we can fix this redundancy assuming that the sum $j_1 \vec{n}_1+j_2 \vec{n}_2$ is in the positive $z$ direction while the vector $\vec{n}_1\times\vec{n}_2$ in the positive $y$ direction. Once chosen this orientation, the four unit-vectors $\vec{n}_1\mdots \vec{n}_4$ (which satisfying the closure condition) depend only on two parameters. These two parameters can be chosen to be the dihedral angle $\cos\theta_{12}=\vec{n}_1\cdot\vec{n}_2$ and the twisting angle $\tan\phi_{(12)(34)}=\frac{(\vec{n}_1\times \vec{n}_2)\cdot(\vec{n}_3\times \vec{n}_4)}{|\vec{n}_1\times \vec{n}_2|\;|\vec{n}_3\times \vec{n}_4|}$\;.}.

 Nodes of the spin network can be labelled with coherent intertwiners. In fact such states provide an overcomplete basis of $\mc{H}_{\Gamma_5}$. Calling $v_i^{m_1\cdots m_4}$ the standard recoupling basis for intertwiners, we can define the coefficients
\begin{equation}
\Phi_i(\vec{n}_1\mdots \vec{n}_4)=v_i^{m_1\cdots m_4} \Phi_{m_1\cdots m_4}(\vec{n}_1\mdots \vec{n}_4)\;.
\end{equation}
We define a \emph{coherent spin network} $|\Gamma_5,j_{ab},\Phi_a\rangle$ as the state labelled by ten spins $j_{ab}$ and $4\times 5$ normals $\vec{n}_{ab}$ and given by the superposition 
\begin{equation}
|\Gamma_5,j_{ab},\Phi_a(\vec{n})\rangle= \sum_{i_1\cdots i_5}\,\Big(\prod_{a=1}^{5} \Phi_{i_a}(\vec{n}_{ab})\Big) |\Gamma_5,j_{ab},i_a\rangle\;.
\end{equation}
The expectation value of the metric operator on a coherent spin network  is simply
\begin{equation}
\langle \Gamma_5,j_{ab},\Phi_a|E_c^{a}\!\cdot\!E_c^{b}|\Gamma_5,j_{ab},\Phi_a\rangle\simeq\gamma^2 j_{ca} j_{cb} \; \vec{n}_{ca}\cdot \vec{n}_{cb}
\end{equation}
in the large spin limit. As a result we can choose the normals $\vec{n}_{ab}$ so that the coherent spin network state is peaked on a given intrinsic geometry of $\Sigma$.

\margin{geometry} Normals in different tetrahedra cannot be chosen independently if we want to peak on a Regge geometry \cite{Regge:1961px}. The relation between normals is provided by the requirement that they are computed from the lengths of the edges of the triangulation $\Delta_5$. \margin{simplify!} In fact, a state with generic normals (satisfying the closure condition (\ref{eq:closure})) is peaked on a discontinuous geometry. This fact can be seen in the following way: let us consider an edge of the triangulation $\Delta_5$; this edge is shared by three tetrahedra; for each tetrahedron we can compute the expectation value of the length operator for an edge in its boundary \cite{Bianchi:2008es}; however in general the expectation value of the length of an edge seen from different tetrahedra will not be the same; this fact shows that the geometry is \emph{discontinuous}. The requirement that the semiclassical state is peaked on a Regge geometry amounts to a number of relations between the labels $\vec{n}_{ab}$. In the case of the boundary of a Euclidean $4$-simplex (excluding the `rectangular' cases discussed in \cite{Barrett:1997tx}), the normals turn out to be completely fixed once we give the areas of the ten triangles or equivalently the ten spins $j_{ab}$,
\begin{equation}
\vec{n}_{ab}=\vec{n}_{ab}(j_{cd})\;.
\label{eq:continuity}
\end{equation}
This assignment of normals guarantees that the geometry we are peaking on is Regge-like. In particular, in this paper we are interested in the case of a $4$-simplex which is approximately regular. In this case the spins labelling the links are of the form $j_{ab}=j_0+\delta j_{ab}$ with $\frac{\delta j_{ab}}{j_0}\ll 1$ and a perturbative expression for the normals solving the continuity condition is available:
\begin{equation}
\vec{n}_{ab}(j_0+\delta j)=\vec{n}_{ab}(j_0)+\sum_{cd}v^{(ab)(cd)} \delta j_{cd}\;.
\end{equation}
The coefficients $v^{(ab)(cd)}$ can be computed in terms of the derivative of the normals $\vec{n}_{ab}$ (for a given choice of orientation, see footnote \ref{footnote:orientation}) with respect to the ten edge lengths, using the Jacobian of the transformation from the ten areas to the ten edge lengths of the 4-simplex. 

\margin{phases} In the following we are interested in superpositions over spins of coherent spin networks. As coherent intertwiners are defined only up to a spin-dependent arbitrary phase, a choice is in order. We make the canonical choice of phases described in \cite{Barrett:2009gg}. We briefly recall it here. Consider a non-degenerate Euclidean 4-simplex; two tetrahedra $t_a$ and $t_b$ are glued at the triangle $f_{ab}\equiv f_{ba}$. Now, two congruent triangles $f_{ab}$ and $f_{ba}$ in $\mbb{R}^3$ can be made to coincide via a unique rotation $R_{ab}\in SO(3)$ which, together with a translation, takes one outward-pointing normal to minus the other one,
\begin{equation}
R_{ab} \vec n_{ab}=-\vec n_{ba}\;.
\end{equation} 
The canonical choice of phase for the spin coherent states $|j_{ab},\vec n_{ab}\rangle$ and $|j_{ab},\vec n_{ba}\rangle$ entering the coherent intertwiners $\Phi_a$ and $\Phi_{b}$ is given by lifting the rotation $R_{ab}$ to a $SU(2)$ transformation $g_{ab}$ and requiring that   
\begin{equation}
|j_{ab},\vec n_{ba}\rangle=D^{(j_{ab})}(g_{ab})\, J |j_{ab},\vec n_{ab}\rangle
\end{equation}
where $J:\mathcal{H}_j\rightarrow \mathcal{H}_j$ is the standard antilinear map for $SU(2)$ representations defined by 
\begin{equation}
\langle \epsilon |\big(|\alpha\rangle \otimes J|\beta\rangle\big)=\langle \beta|\alpha\rangle\quad\text{with}\quad |\alpha\rangle,|\beta\rangle\;\in\,\mathcal{H}_j
\end{equation}
and $\langle \epsilon |$ is the unique intertwiner in $\mathcal{H}_j\otimes \mathcal{H}_j$. In the following we will always work with coherent spin networks $|\Gamma_5,j_{ab},\Phi_a(\vec{n}(j))\rangle$ satisfying the continuity condition, and with the canonical choice for the arbitrary phases of coherent states. From now on we use the shorter notation $|j,\Phi(\vec n)\rangle$.

Coherent spin networks are eigenstates of the diagonal components of the metric operator, namely the area operator for the triangles of $\Delta_5$. \margin{extrinsic} The extrinsic curvature to the manifold $\Sigma$ measures the amount of change of the $4$-normal to $\Sigma$, parallel transporting it along $\Sigma$. In a piecewise-flat context, the extrinsic curvature has support on triangles, that is it is zero everywhere except that on triangles. For a triangle $f_{ab}$, the extrinsic curvature $K_{ab}$ is given by the angle between the $4$-normals $N_a^\mu$ and $N_b^\mu$ to two tetrahedra $t_a$ and $t_b$ sharing the face $f_{ab}$. As the extrinsic curvature is the momentum conjugate to the intrinsic geometry, we have that a semiclassical state cannot be an eigenstate of the area as it would not be peaked on a given extrinsic curvature. In order to define a state peaked both on intrinsic and extrinsic geometry, we consider a superposition of coherent spin networks,
\begin{equation}
|\Psi_0\rangle=\sum_{j_{ab}} \psi_{j_0,\phi_0}(j) |j,\Phi(\vec n)\rangle\;,
\end{equation}
with coefficients $\psi_{j_0,\phi_0}(j)$ given by a gaussian times a phase,
\begin{equation}
\psi_{j_0,\phi_0}(j)=\frac{1}{N}
\exp\Big(-\sum_{ab,cd} \alpha^{(ab)(cd)}\, \frac{j_{ab}-j_{0 ab}}{\sqrt{j_{0 ab}}}\,\frac{j_{cd}-j_{0 cd}}{\sqrt{j_{0 cd}}}\Big)\;\, 
\exp\Big({-i\sum_{ab} \phi_0^{ab}\,(j_{ab}-j_{0 ab})}\Big)\;.
\end{equation}
As we are interested in a boundary configuration peaked on the geometry of a regular $4$-simplex, we choose all the background spins to be equal, $j_{0ab}\equiv j_0$. Later we will consider an asymptotic expansion for large $j_0$. The phases $\phi_0^{ab}$ are also chosen to be equal. The extrinsic curvature at the face $f_{ab}$ in a regular $4$-simplex is $K_{ab}=\arccos N_a\cdot N_b=\arccos(-\frac{1}{4})$. In Ashtekar-Barbero variables $(E_0,A_0)$ we have
\begin{equation}
\phi_0\equiv\phi_0^{ab}=\gamma  K_{ab}= \gamma \arccos(-1/4)\;.
\end{equation} 
The $10\times 10$ matrix $\alpha^{(ab)(cd)}$ is assumed to be complex with positive definite real part. Moreover we require that it has the symmetries of a regular $4$-simplex. We introduce the matrices $P_k^{(ab)(cd)}$ with $k=0,1,2$ defined as 
\begin{align}
P_0^{(ab)(cd)}=1 &\quad \text{if}\quad (ab)=(cd)\quad \text{and zero otherwise},\\
P_1^{(ab)(cd)}=1 &\quad \text{if}\quad  \{a=c,b\neq d\} \;\;\text{or a permutation of it} \quad \text{and zero otherwise},\\
P_2^{(ab)(cd)}=1 &\quad \text{if}\quad (ab)\neq(cd)\quad \text{and zero otherwise}.
\end{align}
Their meaning is simple: a couple $(ab)$ identifies a link of the graph $\Gamma_5$; two links can be either coincident, or touching at a node, or disjoint. The matrices $P_k^{(ab)(cd)}$ correspond to these three different cases. Using the basis $P_k^{(ab)(cd)}$ we can write the matrix $\alpha^{(ab)(cd)}$ as
\begin{equation}
\alpha^{(ab)(cd)}=\sum_{k=0}^{2} \alpha_k\,P_k^{(ab)(cd)}\;.
\label{eq:alpha}
\end{equation}
As a result our ansatz for a semiclassical boundary state $|\Psi_0\rangle$ is labelled by a (large) half-integer $j_0$ and has only three complex free parameters, the numbers $\alpha_k$.

\section{The new spin foam dynamics}\label{sec:W}
The dynamics is implemented in terms of a spin foam functional $\langle W|$. Here we are interested in its components on the Hilbert space spanned by spin networks with graph $\Gamma_5$. 
The sum over two-complexes can be implemented in terms of a formal perturbative expansion in the parameter $\lambda$ of a Group Field Theory \cite{Freidel:2005qe}:
\begin{equation}
\langle W|\Gamma_5,j_{ab},i_a\rangle=\sum_{\sigma}\lambda^{N_\sigma} W(\sigma)\label{eq:vertex expansion}
\end{equation}
The sum is over spinfoams (colored 2-complexes) whose boundary is the spin network $(\Gamma_5,j_{ab},i_a)$, $W(\sigma)$ is the spinfoam amplitude
\begin{equation}
W(\sigma)=\prod_{f\subset\sigma} W_f\prod_{v\subset\sigma} W_v
\end{equation}
where $W_v$ and $W_f$ are the vertex and face amplitude respectively. The quantity $N_\sigma$ in (\ref{eq:vertex expansion}) is the number of vertices in the spin foam $\sigma$, therefore the formal expansion in $\lambda$ is in fact a \emph{vertex expansion}.

 The spin foam models we consider here are the EPRL$_\gamma$ \cite{Engle:2007wy} and FK$_\gamma$ \cite{Freidel:2007py} models. We restrict attention to $0<\gamma<1$; in this case the two models coincide. The vertex amplitude is given by
\begin{equation}
	W_v(j_{ab},i_a)=\sum_{i_a^+\,i_a^-}\, \{15j\}\big(j_{ab}^+,i_a^+\big)\;\, \{15j\}\big(j_{ab}^-,i_a^-\big)\;
	\prod_a  f^{i_a}_{i_a^+ i_a^-}(j_{ab})
	\label{eq:EPRL}
\end{equation}
where the unbalanced spins $j^+, j^-$ are
\begin{equation}
j_{ab}^\pm=\gamma^\pm j_{ab},\quad\quad \gamma^\pm=\frac{1\pm\gamma}{2}\;.
\end{equation}
This relation puts restrictions\footnote{Formula \eqref{eq:EPRL} is well-defined only for $j^\pm$ half-integer. As a result, for a fixed value of $\gamma$, there are restrictions on the boundary spin $j$. For instance, if we choose $\gamma=1/n$ with $n$ integer, then we have that $j$ has to be integer and $j\geq n$, i.e. $j\in\{n,n+1,n+2,\cdots\}$.} on the value of $\gamma$ and of $j_{ab}$. The fusion coefficients $f^{i_a}_{i_a^+ i_a^+}(j_{ab})$ are defined in \cite{Engle:2007wy} (see also \cite{Alesci:2008un}) and built out of the intertwiner $v_i^{m_1\cdots m_4}$ in $\mc{H}_{j_1}\otimes\cdot\cdot\otimes \mc{H}_{j_4}$ and the intertwiners $v_{i_\pm}^{m^\pm_1\cdots m^\pm_4}$ in $\mc{H}_{j^\pm_1}\otimes\cdots\otimes \mc{H}_{j^\pm_4}$. Defining a map $Y:\mc{H}_j\rightarrow \mc{H}_{j^+}\otimes \mc{H}_{j^-}$ with matrix elements $Y^m_{m^+ m^-}=\langle j^+, m^+; j^-, m^-|\, Y |j,m\rangle$ given by Clebsh-Gordan coefficients, we have that the fusion coefficients  $f^{i_a}_{i_a^+ i_a^+}$ are given by
\begin{equation}
f^{i_a}_{i_a^+ i_a^+}= Y_{m_1 m_1^+ m_1^-}\cdots Y_{m_4 m_4^+ m_4^-}\;\;v_i^{m_1\cdots m_4}\;\;v_{i_+}^{m^+_1\cdots m^+_4}\;\;v_{i_-}^{m^-_1\cdots m^-_4}\;.
\end{equation}
Indices are raised and lowered with the Wigner metric.

Throughout this paper we will restrict attention to the lowest order in the vertex expansion. To this order, the boundary amplitude of a spin network state with graph $\Gamma_5$ is given by
\begin{equation}
\langle W|\Gamma_5,j_{ab},i_a\rangle=\mu(j_{ab}) W_v(j_{ab},i_a)\;,
\end{equation}
i.e. it involves a single spin foam vertex. 

The function $\mu$ is defined as $\mu(j)=\prod_{ab} W_{f_{ab}}(j)$. A natural choice for the face amplitude is $W_f(j^+,j^-)=(2j^+ +1)(2j^-+1)=(1-\gamma^2)j^2+2j+1$. Other choices can be considered. We assume that $\mu(\lambda j_{ab})$ scales as $\lambda^p$ for some $p$ for large $\lambda$. We will show in the following that, at the leading order in large $j_0$, the LQG propagator (\ref{eq:G}) is in fact independent from the choice of face amplitude, namely from the function $\mu(j)$.

\section{LQG propagator: integral formula}\label{sec:integral formula}
In this section we define the LQG propagator and then provide an integral formula for it. The dynamical expectation value of an operator $\mc{O}$ on the state $|\Psi_0\rangle$ is defined via the following expression
\begin{equation}
\langle \mc{O}\rangle=\frac{\langle W| \mc{O}|\Psi_0\rangle}{\langle W|\Psi_0\rangle}\;.
\end{equation}
The geometric operator we are interested in is the metric operator $E_n^a\cdot E_n^b$ discussed in section \ref{sec:boundary state}. We focus on the \emph{connected} two-point correlation function $G_{nm}^{abcd}$ on a semiclassical boundary state $|\Psi_0\rangle$. It is defined as
 \begin{equation}
G_{nm}^{abcd}=\langle E_n^a\!\cdot\! E_n^b\; E_m^c\!\cdot\! E_m^d\rangle - \langle E_n^a\!\cdot\! E_n^b\rangle\, \langle E_m^c\!\cdot\! E_m^d\rangle\;.
\label{eq:Gnm}
\end{equation}
We are interested in computing this quantity using the boundary state $|\Psi_0\rangle$ introduced in section \ref{sec:boundary state} and the spin foam dynamics \eqref{eq:EPRL}. This is what we call the \emph{LQG propagator}. As the boundary state is a superposition of coherent spin networks, the LQG propagator involves terms of the form $\langle W|\mc{O}|j,\Phi(\vec n)\rangle$. Its explicit formula is
\begin{equation}\label{propagatorexplicit}
\textstyle G^{abcd}_{nm}=\frac{\sum_{j}\psi(j)\langle W| E_n^a\cdot E_n^b\, E_m^c\cdot E_m^d|j,\Phi(\vec n)\rangle}{\sum_{j}\psi(j)\langle W| j,\Phi(\vec n)\rangle}-
\frac{\sum_{j}\psi(j)\langle W| E_n^a\cdot E_n^b|j,\Phi(\vec n)\rangle}{\sum_{j}\psi(j)\langle W| j,\Phi(\vec n)\rangle}\frac{\sum_{j}\psi(j)\langle W| E_m^c\cdot E_m^d|j,\Phi(\vec n)\rangle}{\sum_{j}\psi(j)\langle W| j,\Phi(\vec n)\rangle}
\end{equation}
In the following two subsections we recall the integral formula for the amplitude of a coherent spin network $\langle W|j,\Phi(\vec n)\rangle$ \cite{Conrady:2008ea,Conrady:2009px},\cite{Barrett:2009gg} and derive analogous integral expressions for the amplitude with metric operator insertions $\langle W|E_n^a\!\cdot\! E_n^b|j,\Phi(\vec n)\rangle$ and $\langle W|E_n^a\!\cdot\! E_n^b\, E_m^c\!\cdot\! E_m^d|j,\Phi(\vec n)\rangle$.

\subsection{Integral formula for the amplitude of a coherent spin network}\label{sec:integral1}
The boundary amplitude of a coherent spin network $|j,\Phi(\vec n)\rangle$ admits an integral representation \cite{Conrady:2008ea,Conrady:2009px},\cite{Barrett:2009gg}. Here we go through its derivation as we will use a similar technique in next section.

The boundary amplitude $\langle W|j,\Phi(\vec n)\rangle$ can be written as an integral over five copies of $SU(2)\times SU(2)$ (with respect to the  Haar measure):
\begin{equation}\label{dynamics}
\langle W|j_{ab},\Phi_a(\vec n)\rangle=\sum_{i_a}\Big(\prod_a \Phi_{i_a}(\vec{n})\Big)\langle W|j_{ab},i_a\rangle=\mu(j)\int \prod_{a=1}^5 \dd g_a^+\dd g_a^-\, \prod_{ab} P^{ab}(g^+,g^-)\;.
\end{equation}
The function $P^{ab}(g^+,g^-)$ is given by
\begin{equation}
P^{ab}(g^+,g^-)=\langle j_{ab},-\vec{n}_{ba}|\,Y^\dag D^{(j^+_{ab})}\big((g_a^+)^{-1} g_b^+\big)\otimes D^{(j^-_{ab})}\big((g_a^-)^{-1} g_b^-\big) Y |j_{ab},\vec{n}_{ab}\rangle\;.
\end{equation}
where the map $Y$ is defined in section \ref{sec:W}. Using the factorization property of spin coherent states,
\begin{equation}
Y|j,\vec{n}\rangle=  |j^+,\vec{n}\rangle\otimes |j^-,\vec{n}\rangle\;,
\end{equation}
we have that the function $P^{ab}(g^+,g^-)$ factorizes as
\begin{equation}
P^{ab}(g^+,g^-)=P^{ab+}(g^+)\, P^{ab-}(g^-)
\end{equation}
with
\begin{equation}
P^{ab\pm}= \langle j_{ab},-\vec{n}_{ba}|D^{(j^\pm_{ab})}\big((g_a^\pm)^{-1} g_b^\pm\big) |j_{ab},\vec{n}_{ab}\rangle=\Big(\langle \frac{1}{2},-\vec{n}_{ba}|(g_a^\pm)^{-1} g_b^\pm |\frac{1}{2},\vec{n}_{ab}\rangle\Big)^{2 j^\pm_{ab}}\;.
\label{eq:Pabpm}
\end{equation}
In the last equality we have used (again) the factorization property of spin coherent states to exponentiate the spin $j_{ab}^\pm$. In the following we will drop the $1/2$ in $|\frac{1}{2},\vec{n}_{ab}\rangle$ and write always $|\vec{n}_{ab}\rangle$ for the coherent state in the fundamental representation.

The final expression we get is
\begin{equation}
\langle W|j,\Phi(\vec n)\rangle=\mu(j)\int \prod_{a=1}^5 \dd g_a^+\dd g_a^-\, e^{S}
\label{eq:e^S}
\end{equation}
where the ``action'' $S$ is given by the sum $S=S^+ + S^-$, with
\begin{equation}
S^{\pm}=\sum_{ab}2j_{ab}^\pm\log\langle -\vec n_{ab}|(g^\pm_a)^{-1}g_b^\pm|\vec n_{ba}\rangle\;.
\label{actionBarrett}
\end{equation}

\subsection{LQG operators as group integral insertions}\label{sec:integral2}
In this section we use a similar technique to derive integral expressions for the expectation value of metric operators. In particular we show that
\begin{equation}
\langle W|E_n^a\!\cdot\! E_n^b|j_{ab},\Phi_a(\vec{n})\rangle=\mu(j)\int \prod_{a=1}^5 \dd g_a^+\dd g_a^-\,\; q_n^{ab}(g^+,g^-)\; e^{S}
\label{eq:<EE>}
\end{equation}
and that
\begin{equation}
\langle W|E_n^a\!\cdot\! E_n^b\;E_m^c\!\cdot\! E_m^d|j_{ab},\Phi_a(\vec{n})\rangle=\mu(j)\int \prod_{a=1}^5 \dd g_a^+\dd g_a^-\,\; q_n^{ab}(g^+,g^-)\;q_m^{cd}(g^+,g^-)\; e^{S}
\label{eq:<EE EE>}
\end{equation}
where we assume\footnote{Similar formulae can be found also in the remaining cases but are not needed for the calculation of the LQG propagator.} $n\neq m$ and $a,b,c,d\neq n,m$. The expression for the insertions $q_n^{ab}(g^+,g^-)$ in the integral is derived below.
%
%

We start focusing on $\langle E_n^a\!\cdot\! E_n^b \rangle$ in the case $a\neq b$. The metric field $(E_a^{b})_i$ acts on a state $|j_{ab},m_{ab}\rangle$ as $\gamma$ times the generator $J_{i}$ of $SU(2)$. As a result we can introduce a quantity $Q^{ab}_i$ defined as 
\begin{equation}
Q^{ab}_i(g^+,g^-)=\langle j_{ab},-\vec{n}_{ba}|\,Y^\dag D^{(j^+_{ab})}\big((g_a^+)^{-1} g_b^+\big)\otimes D^{(j^-_{ab})}\big((g_a^-)^{-1} g_b^-\big) Y (E_a^{b})_i|j_{ab},\vec{n}_{ab}\rangle\;,
\end{equation}
so that 
\begin{equation}
\langle W| E^{na}\!\cdot\! E^{nb}|j,\Phi(\vec n)\rangle=\int \prod_{a=1}^5 \dd g_a^+\dd g_a^-\, \delta^{ij} Q^{na}_i Q^{nb}_j {\prod_{cd}}' P^{cd}(g^+,g^-)\;.
\label{eq:QQP}
\end{equation}
The product $\prod'$ is over couples $(cd)$ different from $(na),(nb)$. Thanks to the invariance properties of the map $Y$, we have that
\begin{equation}
Y J^{ab}_i|j_{ab},m_{ab}\rangle=(J_i^{ab+}+J_i^{ab-})Y|j_{ab},m_{ab}\rangle\;.
\end{equation} 
Thus $Q^{ab}_i$ can be written as 
\begin{equation}
Q^{ab}_i= Q^{ab+}_i\, P^{ab-}+ P^{ab+}\, Q^{ab-}_i
\label{eq:Q=QpPm}
\end{equation}
with 
\begin{equation}
Q^{ab\pm}_i=\gamma\; \langle j^\pm_{ab},-\vec{n}_{ba}|D^{(j^\pm_{ab})}\big((g_a^\pm)^{-1} g_b^\pm\big) J^{ab\pm}_i|j^\pm_{ab},\vec{n}_{ab}\rangle\;.
\end{equation}
Now we show that $Q^{ab\pm}_i$ is given by a function $A^{ab\pm}_i$ linear in the spin $j^\pm_{ab}$, times the quantity $P^{ab\pm}$ defined in (\ref{eq:Pabpm}),
\begin{equation}
Q^{ab\pm}_i= A^{ab\pm}_i\;P^{ab\pm}\;.
\label{eq:Qpm=Apm Ppm}
\end{equation}
The function $A^{ab\pm}_i$ is determined as follows. The generator $J_i^{ab\pm}$ of $SU(2)$ in representation $j_{ab}^\pm$ can be obtained as the derivative
\begin{equation}
i\left.\frac{\partial}{\partial \alpha^i}D^{(j_{ab}^\pm)}\big(h(\alpha)\big)\right|_{\alpha^i=0}=J_i^{ab\pm}
\end{equation}
where the group element $h(\alpha)$ is defined via the canonical parametrization $h(\alpha)=\exp(-i \alpha^i \frac{\sigma_i}{2})$. Therefore, we can write $Q^{ab\pm}_i$ as 
\begin{align}\nonumber
Q^{ab\pm}_i=&\;i\,\gamma\left.\frac{\partial}{\partial \alpha^i} \Big( \langle j^\pm_{ab},-\vec{n}_{ba}|D^{(j^\pm_{ab})}\big((g_a^\pm)^{-1} g_b^\pm\big) D^{(j^\pm_{ab})}\big(h(\alpha)\big)|j^\pm_{ab},\vec{n}_{ab}\rangle\Big)\right|_{\alpha^i=0}\\\nonumber
=&\;i\,\gamma\left.\frac{\partial}{\partial \alpha^i} \Big(\gamma\; \langle -\vec{n}_{ba}|(g_a^\pm)^{-1} g_b^\pm h(\alpha)|\vec{n}_{ab}\rangle\Big)^{2j^\pm_{ab}}\right|_{\alpha^i=0}\\
=&\;\gamma\;j^\pm_{ab}\;\langle -\vec n_{ba}|(g^{\pm}_a)^{-1}g^{\pm}_b \sigma^i|\vec n_{ab}\rangle \big(\langle -\vec n_{ba}|(g^{\pm}_a)^{-1}g^{\pm}_b|\vec n_{ab}\rangle\big)^{2j^\pm_{ab}-1}\;.\label{eq:Qpm=3}
\end{align}
Comparing expression (\ref{eq:Qpm=3}) with (\ref{eq:Qpm=Apm Ppm}) and (\ref{eq:Pabpm}), we find that $A^{na\pm}_i$ is given by\margin{proof$\rightarrow$}
\begin{equation}
A^{na\pm}_i=\gamma j_{na}^{\pm} \frac{\langle -\vec n_{an}|(g^{\pm}_a)^{-1}g^{\pm}_n \sigma^i|\vec n_{na}\rangle}
{\langle -\vec n_{an}|(g^{\pm}_a)^{-1}g^{\pm}_n|\vec n_{na}\rangle}\;.
\label{eq:Apm0}
\end{equation}
A vectorial expression for $A^{na\pm}_i$ can be given, introducing the rotation $R_a^{\pm}=D^{(1)}(g_a^{\pm})$,
\begin{equation}
A^{na\pm}_i=\gamma j_{na}^{\pm}\,(R_n^{\pm})^{-1}\frac{R_n^\pm n_{na}-R_a^\pm n_{an}-i  (R_n^\pm n_{na}\times R_a^\pm n_{an})}{1-(R_a^\pm n_{an})\cdot(R_n^\pm n_{na})}\;.
\label{eq:Apm}
\end{equation}

Thanks to (\ref{eq:Q=QpPm}) and (\ref{eq:Qpm=Apm Ppm}), we have that the expression for $Q^{ab}_i$ simplifies to 
\begin{equation}
Q^{ab}_i= A^{ab}_i\;P^{ab}
\end{equation}
with
\begin{equation}
A^{ab}_i= A^{ab+}_i+A^{ab-}_i\;.
\end{equation}
As a result, equation (\ref{eq:QQP}) reduces to
\begin{equation}
\langle W| E_n^{a}\!\cdot\! E_n^{b}|j,\Phi(\vec n)\rangle=\int \prod_{a=1}^5 \dd g_a^+\dd g_a^-\; \delta^{ij} A^{na}_i A^{nb}_j {\prod_{cd}}\; P^{cd}(g^+,g^-)\;,
\label{eq:AAP}
\end{equation}
which is of the form (\ref{eq:e^S}) with the insertion $A^{na}\cdot A^{nb}$. Therefore, comparing with equation (\ref{eq:<EE>}), we have that
\begin{equation}
q_n^{ab}(g^+,g^-)=A^{na}\cdot A^{nb}\;
\label{eq:qab}
\end{equation}
for $a\neq b$. The case with $a=b$ can be computed using a similar technique but the result is rather simple and expected, thus we just state it
\begin{equation}
q_n^{aa}(g^+,g^-)=\gamma^2 j_{na}(j_{na}+1)\;.
\label{eq:qaa}
\end{equation}

As far as $\langle E_n^{a}\!\cdot\! E_n^{b}\, E_m^{c}\!\cdot\! E_m^{d} \rangle$ a similar result can be found. In particular, for $n\neq m$ and $a,b,c,d\neq n,m$ the result is stated at the beginning of this section, equation (\ref{eq:<EE EE>}), with the same expression for the insertion $q_n^{ab}(g^+,g^-)$ as in equation (\ref{eq:qab}) and equation (\ref{eq:qaa}).

Substituting (\ref{eq:<EE>})-(\ref{eq:<EE EE>}) in \eqref{propagatorexplicit} we obtain a new expression for the propagator in terms of group integrals: 
\begin{equation}
G_{nm}^{abcd}=\frac{\sum_j \mu(j)\psi(j) \int \dd g^{\pm}  \;q_n^{ab}\,q_m^{cd}\, e^S}{\sum_j \mu(j)\psi(j)\int\dd g^{\pm}\,e^S}-
\frac{\sum_j \mu(j) \psi(j) \int \dd g^{\pm}\; q_n^{ab} \,e^S}{\sum_j \mu(j)\psi(j)\int\dd g^{\pm}\,e^S}\frac{\sum_j \mu(j)\psi(j) \int \dd g^{\pm} \,q_m^{cd}\, \,e^S}{\sum_j \mu(j)\psi(j)\int\dd g^{\pm}\,e^S}\;.\label{nondiag-nondiagsums}
\end{equation}
This expression with metric operators written as insertions in an integral is the starting point for the large $j_0$ asymptotic analysis of next section.

\section{LQG propagator: stationary phase approximation}\label{sec:stationary phase}
The correlation function (\ref{nondiag-nondiagsums}) depends on the scale $j_0$ fixed by the boundary state. We are interested in computing its asymptotic expansion for large $j_0$. The technique we use is an (extended) stationary phase approximation of a multiple integral over both spins and group elements. In \ref{sec:total action} we put expression (\ref{nondiag-nondiagsums}) in a form to which this approximation can be applied. Then in \ref{sec:connected} we recall a standard result in asymptotic analysis regarding connected two-point functions and in \ref{sec:critical points}-\ref{sec:Hessian} we apply it to our problem.

\subsection{The total action and the extended integral}\label{sec:total action}
We introduce the ``total action'' defined as $S_{\text{tot}}=\log \psi+ S$ or more explicitly as
\begin{align}\nonumber
S_\text{tot}(j_{ab},g_a^{+},g_a^-)=&\;-\frac{1}{2}\sum_{ab,cd} \alpha^{(ab)(cd)}\, \frac{j_{ab}-j_{0 ab}}{\sqrt{j_{0 ab}}}\,\frac{j_{cd}-j_{0 cd}}{\sqrt{j_{0 cd}}}-i\sum_{ab} \phi_0^{ab}\,(j_{ab}-j_{0 ab})\\
&+S^+(j_{ab},g_a^+)+S^-(j_{ab},g_a^-)\;.
\end{align}
Notice that the action $S^++S^-$ is a homogeneous function of the spins $j_{ab}$ therefore, rescaling the spins $j_{0ab}$ and $j_{ab}$ by an interger $\lambda$ so that $j_{0ab}\rightarrow \lambda j_{0ab}$ and $j_{ab}\rightarrow \lambda j_{ab}$, we have that the total action goes to $S_\text{tot} \rightarrow \lambda S_\text{tot}$. We recall also that $q_n^{ab}\rightarrow \lambda^2 q_n^{ab}$. In the large $\lambda$ limit, the sums over spins in expression (\ref{nondiag-nondiagsums}) can be approximated with integrals over continuous spin variables\footnote{The remainder, i.e. the difference between the sum and the integral, can be estimated via Euler-Maclaurin summation formula. This approximation does not affect any finite order in the computation of the LQG propagator.}:
\begin{equation}
\sum_j \mu \int \dd^5 g^{\pm} \;q_n^{ab} \, e^{\lambda S_{\text{tot}}}=\int \dd^{10}j\, \dd^5 g^{\pm}  \; \mu \,q_n^{ab}\, e^{\lambda S_{\text{tot}}}+O(\lambda^{-N})\qquad \forall N>0\;.
\end{equation}
Moreover, notice that the action, the measure and the insertions in (\ref{nondiag-nondiagsums}) are invariant under a $SO(4)$ symmetry that makes an integration $\dd g^+ \dd g^-$ redundant. We can factor out one $SO(4)$ volume, e.g. putting $g_1^+=g_1^-=1$, so that we end up with an integral over $\dd^4 g^\pm=\textstyle{\prod_{a=2}^5} \dd g_a^+\dd g_a^-$.

As a result we can re-write expression (\ref{nondiag-nondiagsums}) in the following integral form
\begin{equation}
G_{nm}^{abcd}=\lambda^4(\frac{\int \dd^{10}j\, \dd^4 g^{\pm} \mu \;q_n^{ab}\,q_m^{cd}\, e^{\lambda S_{\text{tot}}}}{\int \dd^{10}j\, \dd^4 g^{\pm} \mu\,e^{\lambda S_{\text{tot}}}}-
\frac{\int \dd^{10}j\, \dd^4 g^{\pm} \mu\; q_n^{ab} \,e^{\lambda S_{\text{tot}}}}{\int \dd^{10}j\, \dd^4 g^{\pm} \mu\,e^{\lambda S_{\text{tot}}}}\frac{\int \dd^{10}j\, \dd^4 g^{\pm} \mu \,q_m^{cd}\, \,e^{\lambda S_{\text{tot}}}}{\int \dd^{10}j\, \dd^4 g^{\pm} \mu\,e^{\lambda S_{\text{tot}}}})\;.
\end{equation}
To this expression we can apply the standard result stated in the following section.

\subsection{Asymptotic formula for connected two-point functions}\label{sec:connected}
Consider the integral 
\begin{equation}\label{typeofintegral}
F(\lambda)=\int \dd x \,f(x)\,e^{\lambda S(x)} 
\end{equation}
over a region of $\mathbb R^d$, with $S(x)$ and $f(x)$ smooth complex-valued functions such that the real part of $S$ is negative or vanishing, $\text{Re} S\leq 0$. Assume also that the stationary points $x_0$ of $S$ are isolated so that the Hessian at a stationary point $H=S''(x_0)$  is non-singular, $\det H\neq 0$. Under these hypothesis an asymptotic expansion of the integral $F$ for large $\lambda$ is available: it is an extension of the standard stationary phase approximation that takes into account the fact that the action $S$ is complex \cite{Hormander:1990a}. A key role is played by \emph{critical points}, i.e. stationary points $x_0$ for which the real part of the action vanishes, $\text{Re} S(x_0)=0$. Here we assume that there is a unique critical point. Then the asymptotic expansion of $F(\lambda)$ for large $\lambda$ is given by
\begin{equation}\label{NLOformula}
F(\lambda)=\left(\frac{2\pi}{\lambda}\right)^{\frac{d}{2}}\frac{e^{i \,\text{Ind}\,H} e^{\lambda S(x_0)}}{\sqrt{\left|\det H\right|}}\left( f(x_0)+\frac{1}{\lambda}\big(\frac{1}{2} f''_{ij}(x_0) (H^{-1})^{ij}+D\big)+\mathcal O({\textstyle\frac{1}{\lambda^2}})\right)
\end{equation}
with $f''_{ij}=\partial^2 f/\partial x^i\partial x^j$ and $\text{Ind}\,H$ is the index\footnote{The index is defined in terms of the eigenvalues of $h_k$ of the Hessian as $\text{Ind}\,H=\frac{1}{2}\sum_k \text{arg}(h_k)$ with $-\frac{\pi}{2}\leq \text{arg}(h_k) \leq+\frac{\pi}{2}$.} of the Hessian. The term $D$ does not contain second derivatives of $f$, it contains only\footnote{More explicitly, the term $D$ is given by
\begin{equation}
D=f'_{i}(x_0) R'''_{jkl}(x_0) (H^{-1})^{ij} (H^{-1})^{kl}+\frac{5}{2} f(x_0) R'''_{ijk}(x_0) R'''_{mnl}(x_0) (H^{-1})^{im} (H^{-1})^{jn} (H^{-1})^{kl}
\end{equation}
with $R(x)=S(x)-S(x_0)-\frac{1}{2}H_{ij}(x_0) (x-x_0)^i (x-x_0)^j$.} $f(x_0)$ and $f'_{i}(x_0)$. Now we consider three smooth complex-valued functions $g$, $h$ and $\mu$. A connected 2-point function relative to the insertions $g$ and $h$ and w.r.t. the measure $\mu$ is defined as
\begin{equation}\label{2-point explicit}
G=\frac{\int\dd x\,\mu(x)\,g(x)h(x)\,e^{\lambda S(x)}}{\int\dd x\,\mu(x)\,e^{\lambda S(x)}}-\frac{\int\dd x\,\mu(x)\,g(x)\,e^{\lambda S(x)}}{\int\dd x\,\mu(x)\,e^{\lambda S(x)}} \frac{\int\dd x\,\mu(x)\,h(x)\,e^{\lambda S(x)}}{\int\dd x\,\mu(x)\,e^{\lambda S(x)}}\;.
\end{equation}
Using \eqref{NLOformula} it is straightforward to show that the (leading order) asymptotic formula for the connected 2-point function is simply
\begin{equation}
G=\frac{1}{\lambda}\,(H^{-1})^{ij}\,g'_i(x_0)\,h'_j(x_0)\,+\mathcal O({\textstyle\frac{1}{\lambda^2}})\;.\label{eq:connected formula}
\end{equation}
Notice that both the measure function $\mu$ and the disconnected term $D$ do not appear in the leading term of the connected 2-point function; nevertheless they are present in the higher orders (loop contributions). The reason we are considering the quantity $G$, built from integrals of the type \eqref{typeofintegral}, is that the LQG propagator has exactly this form. Specifically, in sections \ref{sec:critical points} we determine the critical points of the total action, in \ref{sec:Hessian} we compute the Hessian of the total action and the derivative of the insertions evaluated at the critical points, and in \ref{sec:propagator} we state our result.

\subsection{Critical points of the total action}\label{sec:critical points}
The real part of the total action is given by
\begin{align}
\text{Re} S_{\text{tot}}=&\;-\sum_{ab,cd} (\text{Re}\, \alpha)^{(ab)(cd)}\, \frac{j_{ab}-j_{0 ab}}{\sqrt{j_{0 ab}}}\,\frac{j_{cd}-j_{0 cd}}{\sqrt{j_{0 cd}}}+\\
&+\sum_{ab}j_{ab}^- \log\frac{1-(R_a^- n_{ab})\cdot (R_b^- n_{ba})}{2}+\sum_{ab}j_{ab}^+ \log\frac{1-(R_a^+ n_{ab})\cdot (R_b^+ n_{ba})}{2}\;.
\end{align}
Therefore, having assumed that the matrix $\alpha$ in the boundary state has positive definite real part, we have that the real part of the total action is negative or vanishing,  $\text{Re} S_{\text{tot}}\leq 0$. In particular the total action vanishes for the configuration of spins $j_{ab}$ and group elements $g_a^\pm$ satisfying 
\begin{align}
&j_{ab}=j_{0ab}\;,\label{eq:j critical}\\[6pt]
&g_a^\pm\;\; \text{such that}\;\;R_a^\pm n_{ab}(j)=-R_b^\pm n_{ba}(j)\;.
\label{eq:g critical}
\end{align}

Now we study the stationary points of the total action and show that there is a unique stationary point for which $\text{Re} S_{\text{tot}}$ vanishes.

 The analysis of stationary points of the action $S^+ +S^-$ with respect to variations of the group variables $g_a^{\pm}$ has been performed in full detail by Barrett et al. in \cite{Barrett:2009gg}. Here we briefly summarize their result as they apply unchanged to the total action. We invite the reader to look at the original reference for a detailed derivation and a geometrical interpretation of the result.
 
The requirement that the variation of the total action with respect to the group variables $g_a^{\pm}$ vanishes, $\delta_g S_{\text{tot}}=0$, leads to the two sets of equations (respectively for the real and the imaginary part of the variation):
\begin{equation}
\sum_{b\neq a} j_{ab}^\pm \frac{R_a^\pm n_{ab}-R_b^\pm n_{ba}}{1-(R_a^\pm n_{ab})\cdot (R_b^\pm n_{ba})}=0\quad,\quad
\sum_{b\neq a} j_{ab}^\pm \frac{(R_a^\pm n_{ab})\times (R_b^\pm n_{ba})}{1-(R_a^\pm n_{ab})\cdot (R_b^\pm n_{ba})}=0\;.
\label{eq:g stationarity}
\end{equation}
When evaluated at the maximum point (\ref{eq:g critical}), these two sets of equations are trivially satisfied. \margin{proof$\rightarrow$} Infact the normals $\vec{n}_{ab}$ in the boundary state are chosen to satisfy the closure condition (\ref{eq:closure}) at each node. Therefore the critical points in the group variables are given by all the solutions of equation (\ref{eq:g critical}). 

For normals $\vec n_{ab}$ which define non-degenerate tetrahedra and satisfy the continuity condition (\ref{eq:continuity}), the equation $R_a n_{ab}=-R_b n_{ba}$ admits two distinct sets of solutions, up to global rotations. These two sets are related by parity. The two sets can be lifted to $SU(2)$. We call them $\bar{g}^+_a$ and $\bar{g}^-_a$. Out of them, four classes of solutions for the couple $(g^+_a,g^-_a)$ can be found. They are given by
\begin{equation}\label{4classes}
(\bar g_a^+,\bar g_a^-),\;(\bar g_a^-,\bar g_a^+),\;(\bar g_a^+,\bar g_a^+),\;(\bar g_a^-,\bar g_a^-)\;.
\end{equation}
The geometrical interpretation is the following. The couples $(j_{ab}\vec n_{ab},j_{ab}\vec n_{ab})$ are interpreted as the selfdual and anti-selfdual parts (with respect to some ``time" direction, e.g. $(0,0,0,1)$) of area bivectors associated to triangles in 4-dimensions; since these bivectors are diagonal, they live in the 3-dimensional subspace of $\mathbb R^4$ orthogonal to the chosen ``time'' direction. Because of the closure condition (\ref{eq:closure}), for a fixed $n$ the four bivectors $(j_{na} \vec n_{na}, j_{na} \vec n_{na})$ define an embedding of a tetrahedron in $\mathbb R^4$. The two group elements $g_a^+$ and $g_a^-$ of the action \eqref{actionBarrett} define an SO(4) element which rotates the ``initial'' tetrahedron. The system (\ref{eq:g critical}) is a gluing condition between tetrahedra. The first two classes of solutions in (\ref{4classes}) glue five tetrahedra into two Euclidean non-degenerate 4-simplices related by a reflection, while the second two classes correspond to degenerate configurations with the $4$-simplex living in the three-dimensional plane orthogonal to the chosen ``time'' direction. 

The evaluation of the action $S(j_{ab},g^+_a,g^-_a)=S^+(j_{ab},g^+_a)+S^-(j_{ab},g^-_a)$ on the four classes of critical points gives
\begin{align}
S(j_{ab},\bar{g}^+_a,\bar{g}^-_a)=&\,+S_{\text{Regge}}(j_{ab})\;,\\
S(j_{ab},\bar{g}^-_a,\bar{g}^+_a)=&\,-S_{\text{Regge}}(j_{ab})\;,\\
S(j_{ab},\bar{g}^+_a,\bar{g}^+_a)=&\,+\gamma^{-1}S_{\text{Regge}}(j_{ab})\;,\\
S(j_{ab},\bar{g}^-_a,\bar{g}^-_a)=&\,-\gamma^{-1}S_{\text{Regge}}(j_{ab})\;,
\label{eq:S4}
\end{align}
where $S_{\text{Regge}}(j_{ab})$ is Regge action for a single $4$-simplex with triangle areas $A_{ab}=\gamma j_{ab}$ and dihedral angles $\phi_{ab}(j)$ written in terms of the areas
\begin{equation}
S_{\text{Regge}}(j_{ab})=\sum_{ab}  \gamma j_{ab} \phi_{ab}(j)\;.
\end{equation}

Now we focus on stationarity of the total action with respect to variations of the spin labels $j_{ab}$. We fix the group elements $(g_a^+,g_a^-)$ to belong to one of the four classes (\ref{4classes}). For the first class we find
\begin{equation}
0=\left.\frac{\partial S_{\text{tot}}}{\partial j_{ab}}\right|_{(\bar{g}_a^+,\bar{g}_a^-)}=-\sum_{cd}\,\frac{\alpha^{(ab)(cd)}(j_{cd}-j_{0 cd})}{\sqrt{j_{0 ab}}\sqrt{j_{0 cd}}}-i \phi_0^{ab}+i \frac{\partial S_{\text{Regge}}}{\partial j_{ab}}\;.
\label{eq:j stationary}
\end{equation}
The quantity $\partial S_{\text{Regge}}/\partial j_{ab}$ is $\gamma$ times the extrinsic curvature at the triangle $f_{ab}$ of the boundary of a $4$-simplex with triangle areas $A_{ab}=\gamma j_{ab}$. As the phase $\phi_0^{ab}$ in the boundary state is choosen to be exactly $\gamma$ times the extrinsic curvature, we have that equation (\ref{eq:j stationary}) vanishes for $j_{ab}=j_{0ab}$. Notice that, besides being a stationary point, this is also a critical point of the total action as stated in (\ref{eq:j critical}). 

On the other hand, if equation (\ref{eq:j stationary}) is evaluated on group elements belonging to the classes $(\bar{g}_a^-,\bar{g}_a^+)$, $(\bar{g}_a^-,\bar{g}_a^-)$, $(\bar{g}_a^+,\bar{g}_a^+)$, we have that there is no cancellation of phases and therefore no stationary point with respect to variations of spins. This is the feature of the phase of the boundary state: it selects a classical contribution to the asymptotics of a spin foam model, a fact first noticed by Rovelli for the Barrett-Crane model in \cite{Rovelli:2005yj}.

\subsection{Hessian of the total action and derivatives of the insertions}\label{sec:Hessian}
Here we compute the Hessian matrix of the total action $S_\text{tot}$ at the critical point $j_{ab}=j_{0ab}$, $(g_a^+,g_a^-)=(\bar{g}_a^+,\bar{g}_a^-)$. We introduce a local chart of coordinates $(\vec{p}_a^+,\vec{p}_a^-)$ in a neighborhood of the point $(\bar{g}_a^+,\bar{g}_a^-)$ on $SU(2)\times SU(2)$. The parametrization is defined as follows: we introduce 
\begin{equation}
g^\pm_a(p^\pm_a)=h(p^\pm_a)\,\bar{g}_a^\pm
\end{equation}
with $h(p^\pm_a)=\sqrt{1-|\vec{p}_a^\pm|^2}\, + i \vec{p}^\pm_a \cdot\vec\sigma$. The vector $\vec{p}_a^\pm$ is assumed to be in a neighborhood of the origin, which corresponds to the critical point $\bar{g}_a^\pm$. We introduce also the notation $n_a^{\pm}$
\begin{equation}
n_a^\pm=\bar{R}_a^\pm n_a
\end{equation}
where $\bar{R}_a^\pm$ is the rotation associated to the $SU(2)$ group element $\bar{g}_a^\pm$. The bivectors $(j_{ab} n_{ab}^+,j_{ab} n_{ab}^-)$ have the geometrical interpretation of area bivectors associated to the triangles of a $4$-simplex with faces of area proportional to $j_{ab}$.

The Hessian matrix is obtained computing second derivatives of the total action with respect to $j_{ab}$, $p_a^+$ and $p_a^-$, and evaluating it at the point $j_{ab}=j_{0ab}$ and $p_a^\pm=0$. With this definitions we have that the (gauge-fixed) Hessian matrix is a $(10+12+12)\times(10+12+12)$ matrix (as it does not contain derivatives w.r.t. $g^\pm_1$) and has the following structure:
\begin{equation}
S''_\text{tot}=\left(\begin{array}{ccc}
	\frac{\partial^2 S_\text{tot}}{\partial j\partial j}&  
0_{10\times 12}&0_{10\times 12}\\
0_{12\times 10} & \frac{\partial^2 S_\text{tot}}{\partial p^+\partial p^+} & 0_{12\times 12} \\[0.1cm]
0_{12\times 10} & 0_{12\times 12} & \frac{\partial^2 S_\text{tot}}{\partial p^-\partial p^-} 
\end{array}\right)
\label{eq:S''}
\end{equation}
as \margin{proofs}
\begin{equation}
\left.\frac{\partial^2 S_\text{tot}}{\partial p_a^{i\pm}\partial p_b^{j\mp}}\right|_{\vec p=0}\!\!=0\;\;,\quad \left.\frac{\partial^2 S_\text{tot}}{\partial j_{ab} \partial p_c^{j\mp}}\right|_{\vec p=0}\!\!=0\;.
\end{equation}
For the non-vanishing entries we find
\begin{align}
&Q_{(ab)(cd)}=\left.\frac{\partial^2 S_\text{tot}}{\partial  j_{ab}\partial  j_{cd}}\right|_{\vec p=0}\!\!=-\frac{\alpha^{(ab)(cd)}}{\sqrt{j_{0ab}}\sqrt{j_{0cd}}}+(S''_\text{Regge})_{(ab)(cd)}\;,\label{eq:Q}\\
&H^\pm_{(ai)(bj)}=\left.\frac{\partial^2 S_\text{tot}}{\partial p_a^{i\pm}\partial p_b^{j\pm}}\right|_{\vec p=0}\!\!=2 i \gamma^\pm  j_{0ab} (\delta^{ij}-n_{ab}^{i\pm}n_{ab}^{j\pm}+i\epsilon^{ijk}n_{ab}^{k\pm})\;,\\
&H^\pm_{(ai)(aj)}=\left.\frac{\partial^2 S_\text{tot}}{\partial p_a^{i\pm}\partial p_a^{j\pm}}\right|_{\vec p=0}\!\!=-2 i \gamma^\pm \sum_{b\neq a} j_{0ab}(\delta^{ij}-n_{ab}^{i\pm} n_{ab}^{j\pm})\;,
\end{align}
where we have defined the 10$\times$10 matrix of second derivatives of the Regge action
\begin{equation}
(S''_\text{Regge})_{(ab)(cd)}=\left.\frac{\partial^2{S_\text{Regge}}}{\partial j_{ab}\partial j_{cd}}\right|_{j_{0ab}}\;.
\end{equation}
\margin{proofs} We report also the first derivatives of the insertion $q_n^{ab}(g^+,g^-)$ evaluated at the critical point:
\begin{align}
\label{DaAA}
\left.\frac{\partial q_n^{ab}}{\partial \vec p^{\pm}_a} \right|_{\vec p =0}&=i\gamma^2\gamma^\pm j_{0na}j_{0nb}(\vec n_{nb}^\pm-\vec n_{na}\cdot \vec n_{nb} \, \vec n_{na}^\pm +i\, \vec n_{na}^\pm \times\vec n_{nb}^\pm)\;,\\[6pt]
\label{DnAA}
\left.\frac{\partial q_n^{ab}}{\partial \vec p^{\pm}_n}\right|_{\vec p =0}&=-i\gamma^2\gamma^\pm j_{0na}j_{0nb}(\vec n_{na}^\pm+\vec n_{nb}^\pm)(1-\vec n_{na}\cdot \vec n_{nb} )\;,\\[6pt]
\label{DjAA}
\left.\frac{\partial q_n^{ab}}{\partial  j_{cd}}\right|_{\vec p =0}&=\left.\gamma^2\frac{\partial ( j_{na}\vec n_{na}\cdot j_{nb}\vec n_{nb}) }{\partial j_{cd}  }\right|_{j_{0ab}}\;.
\end{align}
We recall that in all these expressions the normals $\vec{n}_{ab}$ are functions of $j_{ab}$ as explained in section \ref{sec:boundary state}. 
These expressions will be used in section \ref{sec:propagator} to compute the leading order of the LQG propagator.

\section{Expectation value of metric operators}\label{sec:expectation values}
Before focusing on the LQG propagator, i.e. on the two-point function, here we briefly discuss the one-point function $\langle E_n^{a}\cdot E_n^{b}\rangle$. Its meaning is the dynamical expectation value of the metric operator. The fact that it is non-vanishing provides the background for the propagator. Using the technique developed in the previous sections we can compute it at the leading order in the large spin expansion. We use the integral formula for the metric operator (\ref{eq:<EE>})-(\ref{eq:<EE EE>}) and the stationary phase analysis of section \ref{sec:stationary phase} and find that the expectation value of the metric operator is simply given by the evaluation of the insertion $q_{n}^{ab}(g^+,g^-)$ at the critical point
\begin{equation}
\langle E_n^{a}\cdot E_n^{b}\rangle=\left.q_{n}^{ab}(g^+,g^-)\right|_{j_{0ab},\bar{g}^+,\bar{g}^-}\;+\mc{O}(j_0).
\end{equation}
For the diagonal components $a=b$ we have that the insertion is simply given by $q_{n}^{aa}=(\gamma j_{na})^{2}$ so that its evaluation at the critical point gives the area square of the triangle $f_{na}$. For the off-diagonal components we have that $q_{n}^{ab}=A_n^a\cdot A_n^b$ where $A_n^{ai}$ is given in equation (\ref{eq:Apm}). Its evaluation at the critical point can be easily found using equation (\ref{eq:g critical}) in expression (\ref{eq:Apm}). We find
\begin{equation}
\left.\vec{A}^{na}\right|_{j_{0ab},\bar{g}^+,\bar{g}^-}=\left.\vec{A}^{na+}\right|_{j_{0ab},\bar{g}^+}+\left.\vec{A}^{na-}\right|_{j_{0ab},\bar{g}^-}=\gamma j_{0na}^+ \vec n_{na}(j_0)+\gamma j_{0na}^-\vec n_{na}(j_0)=\gamma j_{0na} \vec n_{na}(j_0)
\end{equation}
so that $\vec{A}^{na}$ at the critical point evaluates to the classical value $\vec{E}^{a}_{n\text{cl}}=\gamma j_{0na} \vec{n}_{na}(j_0)$, the normal to the face $a$ of the tetrahedron $n$ (normalized to the area of the face). It is the classical counterpart of the operator $(E_n^{a})^i$. Therefore we have that at the leading order the expectation value of the off-diagonal components is given by the dihedral angle between two faces of a tetrahedron
\begin{align}\nonumber
\langle E_n^{a}\cdot E_n^{b}\rangle=&\;\vec{E}^{a}_{n\text{cl}}\cdot\vec{E}^{b}_{n\text{cl}}+\mc{O}(j_0)\\=
&\;\gamma^2 j_{0na} j_{0nb}\, \vec n_{na}(j_0)\cdot \vec n_{nb}(j_0) +\mc{O}(j_0)\;.
\label{eq:expectation1}
\end{align}
 They have the expected geometrical meaning. We observe that the same quantities computed with the Barrett-Crane spinfoam dynamics do not show the right behavior when the off-diagonal components of the metric operator are considered.

Using the same technique we can evaluate the leading order of the two-point function. We have that
\begin{equation}
\langle E_n^{a}\!\cdot\! E_n^{b}\,E_m^{c}\!\cdot\! E_m^{d} \rangle= {E_n^{a}}_\text{cl}\!\cdot\! {E_n^{b}}_\text{cl} \,{E_m^{c}}_\text{cl}\!\cdot\! {E_m^{d}}_\text{cl}+\mathcal{O}(j_0^3)\;.
\label{eq:expectation2}
\end{equation}
The quantity we are specifically interested in in this paper is the \emph{connected} two-point function. It is of order $\mathcal{O}(j_0^3)$, therefore it requires the next-to-leading orders in equations (\ref{eq:expectation1}) and (\ref{eq:expectation2}). Such orders depend on the measure $\mu(j)$. However in the computation of the connected part, these contributions cancel. The technique we use in next section for the calculation of the connected two-point function is the one introduced in section (\ref{sec:connected}) and captures directly the leading order.

\section{LQG propagator: the leading order}\label{sec:propagator}
We have defined the LQG propagator as the connected two-point function $G_{nm}^{abcd}=\langle E_n^a\cdot E_n^b\, E_m^c\cdot E_m^d\rangle-\langle E_n^a\cdot E_n^b\rangle\,\langle E_m^c\cdot E_m^d\rangle$. Using the integral formula (\ref{eq:<EE>})-(\ref{eq:<EE EE>}) and the result (\ref{eq:connected formula}) for the asymptotics of connected two-point functions, we can compute the LQG propagator in terms of (the inverse of) the Hessian of the total action and of the derivative of the metric operator insertions at the critical point. These two ingredients are computed in section \ref{sec:Hessian}. Using them, we find that the LQG propagator is given by
\begin{align}\nonumber
G_{nm}^{abcd}(\alpha)=&\sum_{p,q,r,s} Q^{-1}_{(pq)(rs)}\;\frac{\partial q_n^{ab}}{\partial j_{pq} }\; \frac{\partial q_m^{cd}}{\partial j_{rs}}\;+\\\nonumber
& + \sum_{r,s=2}^5\sum_{i,k=1}^3  \Big((H^+)^{-1}_{(ri)(sk)}\; \frac{\partial q_n^{ab}}{\partial p^{i+ }_r}\;\frac{\partial q_m^{cd}}{\partial p^{k+ }_s} + (H^-)^{-1}_{(ri)(sk)} \frac{\partial q_n^{ab}}{\partial p^{i- }_r} \frac{\partial q_m^{cd}}{\partial p^{k- }_s}\Big)\\[6pt]
&+\mc{O}(j_0^2)
\label{propagatorAtilde}
\end{align}
where all the terms appearing in this expression are defined in section \ref{sec:Hessian}.
%
From this expression we can extract the dependence on the boundary spin $j_0$ and on the Immirzi parameter $\gamma$. We notice that the combinations
\begin{align}
R_{nm}^{abcd}=&\;\frac{1}{\gamma^3 j_0^3}\sum_{p<q,r<s} Q^{-1}_{(pq)(rs)}\;\frac{\partial q_n^{ab}}{\partial j_{pq} }\; \frac{\partial q_m^{cd}}{\partial j_{rs}}\;,\\
X_{nm}^{abcd}=&\;\frac{1}{2\gamma^4 j_0^3}\sum_{r,s=2}^5\sum_{i,k=1}^3  \Big(\frac{1}{\gamma^+}(H^+)^{-1}_{(ri)(sk)}\; \frac{\partial q_n^{ab}}{\partial p^{i+ }_r}\;\frac{\partial q_m^{cd}}{\partial p^{k+ }_s} + \frac{1}{\gamma^-}(H^-)^{-1}_{(ri)(sk)} \frac{\partial q_n^{ab}}{\partial p^{i- }_r} \frac{\partial q_m^{cd}}{\partial p^{k- }_s}\Big)\;,\\
Y_{nm}^{abcd}=&\;\frac{1}{2\gamma^4 j_0^3}\sum_{r,s=2}^5\sum_{i,k=1}^3  \Big(\frac{1}{\gamma^+}(H^+)^{-1}_{(ri)(sk)}\; \frac{\partial q_n^{ab}}{\partial p^{i+ }_r}\;\frac{\partial q_m^{cd}}{\partial p^{k+ }_s} - \frac{1}{\gamma^-}(H^-)^{-1}_{(ri)(sk)} \frac{\partial q_n^{ab}}{\partial p^{i- }_r} \frac{\partial q_m^{cd}}{\partial p^{k- }_s}\Big)\;,
\end{align}
are in fact independent from $j_0$ and from $\gamma$. In terms of these quantities we have that the LQG propagator has the following structure
\begin{align}
G_{nm}^{abcd}(\alpha)=(\gamma j_0)^3 \big(R_{nm}^{abcd}(\alpha)+\gamma X_{nm}^{abcd}+\gamma^2 Y_{nm}^{abcd}\big)\,+\mc{O}(j_0^2)\label{eq:R+X+Y}
\end{align}
where the dependence on $\gamma$ and on $j_0$ has been made explicit now. The matrices $R_{nm}^{abcd}$, $X_{nm}^{abcd}$ and $Y_{nm}^{abcd}$ can be evaluated algebraically. Only the matrix $R_{nm}^{abcd}$ depends on the three parameters $\alpha_0,\alpha_1, \alpha_2$ appearing on the boundary state. We find that 
\begin{equation}
R_{nm}^{abcd}=\left(
\begin{array}{ccc}
	\left(
\begin{array}{ccc}
c_1 &c_3&c_3\\
c_3 &c_2&c_4\\
c_3 &c_4&c_2
\end{array}
	\right) & 
	\left(
\begin{array}{ccc}
c_3 &c_5 &c_6 \\
c_5 &c_3 &c_6 \\
c_6 &c_6 &c_4
\end{array}
	\right) &
	\left(
\begin{array}{ccc}
c_3 &c_6 &c_5 \\
c_6 &c_4 &c_6 \\
c_5 &c_6 &c_3
\end{array}
	\right)\\
\left(
\begin{array}{ccc}
c_3 &c_5 &c_6 \\
c_5 &c_3 &c_6 \\
c_6 &c_6 &c_4
\end{array}
	\right) & 
	\left(
\begin{array}{ccc}
c_2 &c_3 &c_4 \\
c_3 &c_1 &c_3 \\
c_4 &c_3 &c_2
\end{array}
	\right) &
	\left(
\begin{array}{ccc}
c_4 &c_6 &c_6 \\
c_6 &c_3 &c_5 \\
c_6 &c_5 &c_3
\end{array}
	\right)\\
\left(
\begin{array}{ccc}
c_3 &c_6 &c_5 \\
c_6 &c_4 &c_6 \\
c_5 &c_6 &c_3
\end{array}
	\right) & 
	\left(
\begin{array}{ccc}
c_4 &c_6 &c_6 \\
c_6 &c_3 &c_5 \\
c_6 &c_5 &c_3
\end{array}
	\right) &
	\left(
\begin{array}{ccc}
c_2 &c_4 &c_3 \\
c_4 &c_2 &c_3 \\
c_3 &c_3 &c_1
\end{array}
	\right)
\end{array}
	\right)
\end{equation}
where
\begin{align}
c_1=&\;4 \beta_1\;\;,\;\;c_2=4 \beta_2\;\;,\;\;c_3=-\frac{2}{3} \left(2 \beta_0-3 \beta_1+3 \beta_2\right)\;,\\
c_4=&\;\frac{1}{3} \left(8 \beta_0-12 \beta_1\right) \;,\;
c_5=\frac{1}{9} \left(49 \beta_0-93 \beta_1+48 \beta_2\right)\;,\;
c_6=-\frac{1}{9} \left(23
   \beta_0-42 \beta_1+15
   \beta_2\right)\;,
\end{align}
and\footnote{We notice that the inverse of the matrix $Q_{(ab)(cd)}$ can be written in terms of the parameters $\beta_k$ using the formalism (\ref{eq:alpha}) introduced in section \ref{sec:boundary state},
\begin{equation}
(Q^{-1})^{(ab)(cd)}=\sum_{k=0}^2 j_0 \beta_k  P_k^{(ab)(cd)}\;.
\end{equation}
The matrix $Q_{(ab)(cd)}$ is defined in equation (\ref{eq:Q}) and is given by
\begin{equation}
Q^{(ab)(cd)}=\frac{1}{j_0}\sum_{k=0}^2(i h_k-\alpha_k) P_k^{(ab)(cd)}\;,
\end{equation}
with $h_0=-\frac{9}{4}\sqrt{\frac{3}{5}}\;$, $h_1=\frac{7}{8} \sqrt{\frac{3}{5}}\;$, $h_2=-\sqrt{\frac{3}{5}}\;$. 
} 
\begin{align}
\beta_0=&\;\frac{1}{10} \left(-\frac{1}{\alpha_0+6 \alpha_1+3 \alpha _2}+\frac{32}{-8 \alpha_0-8 \alpha_1+16 \alpha_2+i \sqrt{15}}-\frac{5}{\alpha_0-2
   \alpha_1+\alpha_2+i \sqrt{15}}\right)\;,\\
\beta_1=&\;\frac{1}{30}  \left(-\frac{3}{\alpha_0+6 \alpha_1+3 \alpha_2}+\frac{16}{-8 \alpha_0-8 \alpha_1+16 \alpha_2+i \sqrt{15}}+\frac{5}{\alpha_0-2
   \alpha_1+\alpha_2+i \sqrt{15}}\right)\;,\\
\beta_2=&\;\frac{1}{30} \left(-\frac{3}{\alpha_0+6 \alpha_1+3 \alpha_2}-\frac{64}{-8 \alpha_0-8 \alpha_1+16 \alpha_2+i \sqrt{15}}-\frac{5}{\alpha_0-2
   \alpha_1+\alpha_2+i \sqrt{15}}\right)\;.
\end{align}
The matrices $X_{nm}^{abcd}$ and $Y_{nm}^{abcd}$ turn out to be proportional 
\begin{equation}
X_{nm}^{abcd}= \frac{7}{36} Z_{nm}^{abcd} \quad , \quad  Y_{nm}^{abcd}=-i \frac{\sqrt{15}}{36} Z_{nm}^{abcd}\;,
\end{equation}
with the matrix $Z_{nm}^{abcd}$ given by
\begin{equation}
Z_{nm}^{abcd}=\left(
\begin{array}{ccc}
\left(
\begin{array}{ccc}
0 &0&0\\
0 &0&0\\
0 &0&0
\end{array}
	\right) & 
	\left(
\begin{array}{ccc}
0 & -1 & e^{i\frac{\pi}{3}}\\
-1 &0&e^{-i\frac{\pi}{3}}\\
e^{i\frac{\pi}{3}} &e^{-i\frac{\pi}{3}}&0
\end{array}
	\right) &
	\left(
\begin{array}{ccc}
0 & e^{-i\frac{\pi}{3}} & -1\\
e^{-i\frac{\pi}{3}} &0&e^{i\frac{\pi}{3}}\\
-1 &e^{i\frac{\pi}{3}}&0
\end{array}
	\right)\\
	\left(
\begin{array}{ccc}
0 & -1 & e^{i\frac{\pi}{3}}\\
-1 &0&e^{-i\frac{\pi}{3}}\\
e^{i\frac{\pi}{3}} &e^{-i\frac{\pi}{3}}&0
\end{array}
	\right)  & 
	\left(
\begin{array}{ccc}
0 &0&0\\
0 &0&0\\
0 &0&0
\end{array}
	\right) &
	\left(
\begin{array}{ccc}
0 & e^{i\frac{\pi}{3}} & e^{-i\frac{\pi}{3}}\\
e^{i\frac{\pi}{3}} &0&-1\\
e^{-i\frac{\pi}{3}} &-1&0
\end{array}
	\right)\\
	\left(
\begin{array}{ccc}
0 & e^{-i\frac{\pi}{3}} & -1\\
e^{-i\frac{\pi}{3}} &0&e^{i\frac{\pi}{3}}\\
-1 &e^{i\frac{\pi}{3}}&0
\end{array}
	\right) & 
	\left(
\begin{array}{ccc}
0 & e^{i\frac{\pi}{3}} & e^{-i\frac{\pi}{3}}\\
e^{i\frac{\pi}{3}} &0&-1\\
e^{-i\frac{\pi}{3}} &-1&0
\end{array}
	\right) &
	\left(
\begin{array}{ccc}
0 &0&0\\
0 &0&0\\
0 &0&0
\end{array}
	\right)
\end{array}
	\right)\;.
\end{equation}

\bigskip

This is the main result of the paper, the scaling and the tensorial structure of metric correlations in LQG. In the following we collect some remarks on this result:
\begin{itemize}
  \item[-] The LQG propagator scales as $j_0^3$, as expected for correlations of objects with dimensions of area square, $E_n^a\cdot E_n^b\sim (\gamma j_0)^2$.
  \item[-] The off-diagonal components are not suppressed as happened for the Barrett-Crane model \cite{Alesci:2007tx,Alesci:2007tg} and have the same scaling as the diagonal ones.
	\item[-] The contribution $R_{nm}^{abcd}$ in (\ref{eq:R+X+Y}) matches exactly with the matrix of correlations of areas and angles computed in perturbative quantum Regge calculus with a boundary state as done in \cite{Bianchi:2007vf}. 
	\item[-] On the other hand, the `$\gamma$-terms' in (\ref{eq:R+X+Y}), $\gamma X_{nm}^{abcd}+ \gamma^2 Y_{nm}^{abcd}$, are new and proper of the spin foam model. They come from $SU(2)\times SU(2)$ ``group'' fluctuations. They don't contribute to area-area correlations, nor to area-angle correlations. On the other hand, their contribution to angle-angle correlations is non-trivial.
	\item[-] In the limit $\gamma\rightarrow 0$ and  $j_0\rightarrow\infty$ with $\gamma j_0=\text{const}=A_0$, only the Regge contribution survives. It is interesting to notice that the same limit was considered in \cite{Bojowald} in the context of loop quantum cosmology.
	\item[-] The `$\gamma$-terms' have an interesting feature that we now describe. Let us focus on the tensorial components $G_4=G_{12}^{(34)(45)}$ and $G_5=G_{12}^{(35)(45)}$. They are related by a permutation of the vertices $4$ and $5$, keeping the other three vertices fixed. The `Regge-term' is invariant under this permutation, $R_{12}^{(34)(45)}=R_{12}^{(35)(45)}$. On the other hand the `$\gamma$-terms' are not. In particular we have that
\begin{equation}
\gamma X_{12}^{(35)(45)} + \gamma^2 Y_{12}^{(35)(45)}= e^{i\frac{2\pi}{3}} \Big(\gamma X_{12}^{(34)(45)} + \gamma^2 Y_{12}^{(34)(45)}\Big)\;.
\end{equation}
It would be interesting to identify the origin of the phase $\frac{2\pi}{3}$. We notice that the permutation of the vertex $4$ with the vertex $5$ of the boundary spin network corresponds to a parity transformation of the four-simplex. In this sense the `$\gamma$-terms' are parity violating. 
\end{itemize}

In next section we investigate the relation of the result found  with the graviton propagator computed in perturbative quantum field theory.


\section{Comparison with perturbative quantum gravity}\label{sec:perturbative}
The motivation for studying the LQG propagator comes from the fact that it probes a regime of the theory where predictions can be compared to the ones obtained perturbatively in a quantum field theory of gravitons on flat space \cite{Veltman:1975vx,Donoghue:1994dn,Burgess:2003jk}. Therefore it is interesting to investigate this relation already at the preliminary level of a single spin foam vertex studied in this paper. In this section we investigate this relation within the setting discussed in \cite{Alesci:2007tx,Alesci:2007tg,Alesci:2008ff}.

In perturbative quantum gravity\footnote{Here we consider the Euclidean case.}, the graviton propagator in the harmonic gauge is given by
\begin{equation}
\langle h_{\mu\nu}(x) h_{\rho\sigma}(y)\rangle =\frac{-1}{2|x-y|^2}(\delta_{\mu\rho}\delta_{\nu\sigma}+\delta_{\mu\sigma}\delta_{\nu\rho}-\delta_{\mu\nu}\delta_{\rho\sigma}).
\label{eq:feynman prop}
\end{equation}
Correlations of geometrical quantities can be computed perturbatively in terms of the graviton propagator. For instance the angle at a point $x_n$ between two intersecting surfaces $f_{na}$ and $f_{nb}$ is given by\footnote{We thank E.~Alesci for a discussion on this point.}
\begin{equation}
q^{ab}_n=g_{\mu\nu}(x_n)g_{\rho\sigma}(x_n) B_{na}^{\mu\rho}(x_n) B_{nb}^{\nu\sigma}(x_n)
\end{equation} 
where $B^{\mu\nu}$ is the bivector associated to the surface\footnote{To be more specific, we consider local coordinates $(\sigma^1,\sigma^2)$ for a surface $t$ and call $t^{\mu}(\sigma)$ its embedding in the $4d$ manifold. The bivector $B_t^{\mu\nu}(x)$ is defined as
\begin{equation}
B_t^{\mu\nu}(x)=\frac{\partial t^\mu}{\partial\sigma^\alpha}\frac{\partial t^\nu}{\partial\sigma^\beta}\eps^{\alpha\beta}\;.
\end{equation}}.
As a result the angle fluctuation can be written in terms of the graviton field
\begin{equation}
\delta q_n^{ab}= h_{\mu\nu}(x_n) \, (T_n^{ab})^{\mu\nu}\;,
\end{equation}
where we have defined the tensor $(T_n^{ab})^{\mu\nu}= 2 \delta_{\rho\sigma} B_{na}^{\mu\rho}(x_n) B_{nb}^{\nu\sigma}(x_n)$. The angle correlation $(G_{nm}^{abcd})_{\text{qft}}$ is simply given by
\begin{equation}
(G_{nm}^{abcd})_{\text{qft}}=\langle h_{\mu\nu}(x_n) h_{\rho\sigma}(x_m)\rangle\; (T_n^{ab})^{\mu\nu} (T_m^{cd})^{\rho\sigma}\;.
\end{equation}
In particular, this quantity can be computed for couples of surfaces identified by triangles of area $A_0$ living on the boundary of a regular Euclidean $4$-simplex. This quantity has been computed in \cite{Alesci:2007tg} and we report it here for reference, 
\begin{equation*}
\Big(G_{nm}^{abcd}\Big)_{\text{qft}}=\frac{-A_0^3}{18\sqrt{3}\times 512}\left(
\begin{array}{ccc}
\!\!\!\!\!\left(
\begin{array}{ccc}
-16 &6&6\\
6 &-28&16\\
6 &16&-28
\end{array}
	\right)\!\!\! & 
	\left(
\begin{array}{ccc}
6 &4&-7\\
4 &6&-7\\
-7 &-7&16
\end{array}
	\right) &
	\left(
\begin{array}{ccc}
6 &-7&4\\
-7 &16&-7\\
4 &-7&6
\end{array}
	\right)\\
	\left(
\begin{array}{ccc}
6 &4&-7\\
4 &6&-7\\
-7 &-7&16
\end{array}
	\right) & 
\!\!\!	\left(
\begin{array}{ccc}
-28 &6&16\\
6 &-16&6\\
16 &6&-28
\end{array}
	\right)\!\!\! &
	\left(
\begin{array}{ccc}
16 &-7&-7\\
-7 &6&4\\
-7 &4&6
\end{array}
	\right)\\
	\left(
\begin{array}{ccc}
6 &-7&4\\
-7 &16&-7\\
4 &-7&6
\end{array}
	\right) & 
	\left(
\begin{array}{ccc}
16 &-7&-7\\
-7 &6&4\\
-7 &4&6
\end{array}
	\right) &
\!\!\!	\left(
\begin{array}{ccc}
-28 &16&6\\
16 &-28&6\\
6 &6&-16
\end{array}
	\right)\!\!\!\!\!
\end{array}
	\right)
\end{equation*}
The question we want to answer here is if the quantity $(G_{nm}^{abcd})_{\text{qft}}$ and the leading order of the LQG propagator given by equation (\ref{eq:R+X+Y}) can match. As we can identify $\gamma j_0$ with the area $A_0$, we have that the two have the same scaling. The non-trivial part of the matching is the tensorial structure. Despite the fact that we have $9\times 9$ tensorial components, only six of them are independent as the others are related by symmetries of the configuration we are considering. On the other hand the semiclassical boundary state $|\Psi_0\rangle$ we used in the LQG calculation has only three free parameters, $\alpha_0, \alpha_1, \alpha_2$. Therefore we can ask if there is a choice of these $3$ parameters such that we can satisfy the $6$ independent equations given by the matching condition 
\begin{equation}
\Big(G_{nm}^{abcd}(\alpha)\Big)_{\text{lqg}}=\Big(G_{nm}^{abcd}\Big)_{\text{qft}}\;.
\label{eq:matching}
\end{equation}
We find that a solution in terms of the parameters $\alpha_k$ can be found only in the limit of vanishing Immirzi parameter, keeping constant the product $\gamma j_0=A_0$. In this limit we find a unique solution for $\alpha_k$ given by
\begin{align}
\alpha_0=&\;\frac{1}{100}(495616
   \sqrt{3}-45 \sqrt{15}\,i)\;,\\[6pt]
\alpha_1=&\;\frac{1}{200}(-299008 \sqrt{3}+35\sqrt{15}\,i)\;,\\[6pt]
\alpha_2=&\;\frac{1}{25}(31744 \sqrt{3}-5\sqrt{15}\,i)\;.
\end{align}
Therefore the matching condition (\ref{eq:matching}) can be satisfied, at least in the specific limit considered. Having found a non-trivial solution, it is interesting to study the real part of the matrix $\alpha^{(ab)(cd)}$ in order to determine if it is positive definite. Its eigenvalues (with the associated degeneracy) are 
\begin{align}
\lambda_5=&\;9216\sqrt{3}\;,\quad\text{deg}=5\;,\\[6pt]
\lambda_4=&\;\frac{4608\sqrt{3}}{5}\;,\quad\text{deg}=4\;,\\[6pt]
\lambda_1=&\;-\frac{1024\sqrt{3}}{5}\;,\quad\text{deg}=1\;.
\end{align}
We notice that all the eigenvalues are positive except one, $\lambda_1$. The corresponding eigenvector represents conformal rescalings of the boundary state, $j_{0ab}\rightarrow \lambda j_{0ab}$. It would be interesting to determine its origin and to understand how the result depends on the choice of gauge made for the graviton propagator (\ref{eq:feynman prop}).

\section{Conclusions}\label{sec:conclusions}

In this paper we have studied correlation functions of metric operators in loop quantum gravity. The analysis presented involves two distinct ingredients:
\begin{itemize}
	\item The first is a setting for defining correlation functions. The setting is the boundary amplitude formalism. It involves a boundary semiclassical state $|\Psi_0\rangle$ which identifies the regime of interest, loop quantum gravity operators $E_n^a \cdot E_n^b$ which probe the quantum geometry on the boundary, a spin foam model $\langle W|$ which implements the dynamics. The formalism allows to define semiclassical correlation functions in a background-independent context.
	\item The second ingredient consists in an approximation scheme applied to the quantity defined above. It involves a vertex expansion and a large spin expansion. It allows to estimate the correlation functions explicitly. The explicit result can then be compared to the graviton propagator of perturbative quantum gravity. In this paper we focused on the lowest order in the vertex expansion and the leading order in the large spin expansion.
\end{itemize}
The results found in the paper can be summarized as follows:
\begin{itemize}
	\item[1.] In section \ref{sec:boundary state} we have introduced a semiclassical state $|\Psi_0\rangle$ peaked on the intrinsic and the extrinsic geometry of the boundary of a regular Euclidean $4$-simplex. The technique used to build this state is the following: (i) we use the coherent intertwiners introduced in \cite{Livine:2007vk,Conrady:2009px} to define coherent spin networks as in \cite{Barrett:2009gg}; (ii) we choose the normals labelling intertwiners so that they are compatible with a simplicial 3-geometry (\ref{eq:continuity}). This addresses the issue of discontinuous lengths identified in \cite{Bianchi:2008es}; (iii) then we take a gaussian superposition over coherent spin networks in order to peak on extrinsic curvature as in \cite{Rovelli:2005yj,Bianchi:2006uf}. This state is an improvement of the ansatz used in \cite{Alesci:2007tx,Alesci:2007tg,Alesci:2008ff}, as it depends only on the three free parameters $\alpha_0, \alpha_1, \alpha_2$.
	\item[2.] In section \ref{sec:integral formula} we have defined expectation values of geometric observables on a semiclassical state. The LQG propagator is defined in equation (\ref{eq:Gnm}) as a connected correlation function for the product of two metric operators
	\begin{equation}
G_{nm}^{abcd}=\langle E_n^a\!\cdot\! E_n^b\; E_m^c\!\cdot\! E_m^d\rangle - \langle E_n^a\!\cdot\! E_n^b\rangle\, \langle E_m^c\!\cdot\! E_m^d\rangle\;.
\end{equation}
	This is the object that in principle can be compared to the graviton propagator on flat space: the background is coded in the expectation value of the geometric operators and the propagator measures correlations of fluctuations over this background.
	\item[3.] In section \ref{sec:integral2} we have introduced a technique which allows to write LQG metric operators as insertions in a $SO(4)$ group integral. It can be interpreted as the covariant version of the LQG operators. The formalism works for arbitrary fixed triangulation. Having an integral formula for expectation values and correlations of metric operators allows to formulate the large spin expansion as a stationary phase approximation. The problem is studied in detail restricting attention to the lowest order in the vertex expansion, i.e. at the single-vertex level.
	\item[4.] The analysis of the large spin asymptotics is performed in section \ref{sec:stationary phase}. The technique used is the one introduced by Barrett et al in \cite{Barrett:2009gg}. There, the large spin asymptotics of the boundary amplitude of a coherent spin network is studied and four distinct critical points are found to contribute to the asymptotics. Two of them are related to different orientations of a $4$-simplex. The other two come from selfdual configurations. Here, our boundary state is peaked also on extrinsic curvature. The feature of this boundary state is that it selects only one of the critical points, extracting $\exp i S_{\text{Regge}}$ from the asymptotics of the EPRL spin foam vertex. This is a realization of the mechanism first identified by Rovelli in \cite{Rovelli:2005yj} for the Barrett-Crane model. 	
	\item[6.] In \ref{sec:expectation values} we compute expectation values of LQG metric operators at leading order and find that they reproduce the intrinsic geometry of the boundary of a regular $4$-simplex.
	\item[7.] Computing correlations of geometric operators requires going beyond the leading order in the large spin expansion. In section \ref{sec:connected} we derive a formula for computing directly the \emph{connected} two-point correlation function to the lowest non-trivial order in the large spin expansion. The formula is used in section \ref{sec:Hessian}.
	\item[8.] The result of the calculation, the LQG propagator, is presented in section \ref{sec:propagator}. We find that the result is the sum of two terms: a ``Regge term'' and a ``$\gamma$-term''. The Regge term coincides with the correlations of areas and angles computed in Regge calculus with a boundary state \cite{Bianchi:2007vf}. It comes from correlations of fluctuations of the spin variables and depends on the parameters $\alpha_k$ of the boundary state. The ``$\gamma$-term'' comes from fluctuations of the $SO(4)$ group variables. 
	 An explicit algebraic calculation of the tensorial components of the LQG propagator is presented.
	\item[9.] The LQG propagator can be compared to the graviton propagator. This is done in section \ref{sec:perturbative}. We find that the LQG propagator has the correct scaling behaviour. The three parameters $\alpha_k$ appearing in the semiclassical boundary state can be chosen so that the tensorial structure of the LQG propagator matches with the one of the graviton propagator. The matching is obtained in the limit $\gamma\rightarrow 0$ with $\gamma j_0$ fixed. 
\end{itemize}
Now we would like to put these results in perspective with respect to the problem of extracting the low energy regime of loop quantum gravity and spin foams (see in particular \cite{ILQGS:5May2009}). 

Deriving the LQG propagator at the level of a single spin foam vertex is certainly only a first step. Within the setting of a vertex expansion, an analysis of the LQG propagator for a finite number of spinfoam vertices is needed. Some of the techniques developed in this paper generalize to this more general case. In particular superpositions of coherent spin networks can be used to build semiclassical states peaked on the intrinsic and the extrinsic curvature of an arbitrary boundary Regge geometry. Moreover, the expression of the LQG metric operator in terms of $SO(4)$ group integrals presented in this paper works for an arbitrary number of spin foam vertices and allows to derive an integral representation of the LQG propagator in the general case, analogous to the one of \cite{Conrady:2008ea} but with non-trivial insertions. This representation is the appropriate one for the analysis of the large spin asymptotics along the lines discussed for Regge calculus in \cite{Bianchi:2008ae}. The non-trivial question which needs to be answered then is if the semiclassical boundary state is able to enforce semiclassicality in the bulk. Another feature identified in this paper which appears to be general is that, besides the expected Regge contribution, correlations of LQG metric operators have a non-Regge contribution which is proper of the spin foam model. It would be interesting to investigate if this contribution propagates when more than a single spin foam vertex is considered.
%
%
%
%
%
%
%
%

\section*{Acknowledgments}
We thank Carlo Rovelli for many useful discussions. We also thank Emanuele Alesci, John W. Barrett, Florian Conrady, Richard J. Dowdall, Winston J. Fairbairn, Henrique Gomes, Frank Hellmann, Leonardo Modesto, Roberto Pereira and Alejandro Satz for comments and suggestions. E.B. gratefully acknowledges support from Fondazione A.~Della~Riccia.



\providecommand{\href}[2]{#2}\begingroup\raggedright\endgroup

\end{document}